\documentstyle[12pt]{article}

\begin{document}

\baselineskip 6mm
\renewcommand{\thefootnote}{\fnsymbol{footnote}}

\newcommand{\nc}{\newcommand}
\newcommand{\rnc}{\renewcommand}


\rnc{\baselinestretch}{1.24}    
\setlength{\jot}{6pt}       
\rnc{\arraystretch}{1.24}   

\makeatletter
\rnc{\theequation}{\thesection.\arabic{equation}}
\@addtoreset{equation}{section}
\makeatother



\nc{\be}{\begin{equation}}

\nc{\ee}{\end{equation}}

\nc{\bea}{\begin{eqnarray}}

\nc{\eea}{\end{eqnarray}}

\nc{\ben}{\begin{eqnarray*}}

\nc{\een}{\end{eqnarray*}}

\nc{\xx}{\nonumber\\}

\nc{\ct}{\cite}

\nc{\la}{\label}

\nc{\eq}[1]{(\ref{#1})}

\nc{\newcaption}[1]{\centerline{\parbox{6in}{\caption{#1}}}}

\nc{\fig}[3]{

\begin{figure}
\centerline{\epsfxsize=#1\epsfbox{#2.eps}}
\newcaption{#3. \label{#2}}
\end{figure}
}


\def\IR{{\hbox{{\rm I}\kern-.2em\hbox{\rm R}}}}
\def\IB{{\hbox{{\rm I}\kern-.2em\hbox{\rm B}}}}
\def\IN{{\hbox{{\rm I}\kern-.2em\hbox{\rm N}}}}
\def\IC{\,\,{\hbox{{\rm I}\kern-.59em\hbox{\bf C}}}}
\def\IZ{{\hbox{{\rm Z}\kern-.4em\hbox{\rm Z}}}}
\def\IP{{\hbox{{\rm I}\kern-.2em\hbox{\rm P}}}}
\def\IH{{\hbox{{\rm I}\kern-.4em\hbox{\rm H}}}}
\def\ID{{\hbox{{\rm I}\kern-.2em\hbox{\rm D}}}}


\def\Tr{{\rm Tr}\,}
\def\det{{\rm det}}


\def\vare{\varepsilon}
\def\barz{\bar{z}}
\def\barw{\bar{w}}

\begin{titlepage}
\hfill\parbox{5cm} 
{IP/BBSR/2003-34 \\ hep-th/0310137}\\ 
\vspace{25mm}
\begin{center}
{\Large {\bf $D=2,\, {\cal N}=2,$ Supersymmetric theories\\
on Non(anti)commutative Superspace}  }

\vspace{15mm}
B. Chandrasekhar\footnote{chandra@iopb.res.in} and 
Alok Kumar\footnote{kumar@iopb.res.in}
\\[3mm]
{\sl Institute of Physics, Bhubaneswar 751 005, INDIA} \\
\end{center}

\thispagestyle{empty}

\vskip2cm


\centerline{\bf ABSTRACT}
\vskip 4mm
\noindent
The classical action of a two dimensional $ {\cal N}=2$ supersymmetric 
theory, characterized by a general K\"{a}hler potential, is written down
on a non(anti)commutative superspace. 
The action 
has a power series expansion in terms of the determinant of the 
non(anti)commutativity parameter $\,C^{\alpha\beta}$. 
The theory
is explicitly shown to preserve half of the $ {\cal N}=2$ supersymmetry,
to all orders in $\,(\det\,C)^n$.
The results are further generalized to include 
arbitrary superpotentials as well.
\\
\vspace{2cm}

\today

\end{titlepage}

\renewcommand{\thefootnote}{\arabic{footnote}}
\setcounter{footnote}{0}

\tableofcontents
\section{Introduction}

Field theories on noncommutative spaces have been studied 
extensively in the
past few years (for a lucid review and full list of references, 
see~\cite{Douglas:2001ba}), 
even more so, after their appearance in certain limits 
of String and M-theories~\cite{Connes:1997cr,Seiberg:1999vs,Douglas:1997fm}. 
Noncommutativity of the bosonic space-time
coordinates emerges on a  $D$-brane worldvolume, when  a constant
NS-NS two form field is turned on. The study of classical, as well as
the quantum aspects of noncommutative field theories has thrown in
many surprises, both at the 
perturbative 
and nonperturbative level.
The most notable ones being, 
UV/IR mixing~\cite{Minwalla:1999px,VanRaamsdonk:2000rr}, 
noncommutative solitons~\cite{Gopakumar:2000zd}, 
quantum Hall fluid~\cite{Susskind:2001fb} ~etc..

Considering the emergence of such interesting concepts, 
it is natural to seek a generalization of the idea of 
noncommutativity
to be applicable 
to wider situations. 
An obvious thing to check is the possibility of having
non(anti)commutativity between
the superspace coordinates. Indeed, in the context of  
Dijkgraaf-Vafa correspondence~\cite{Dijkgraaf:2002dh}
relating ${\cal N} =1$ supersymmetric gauge theories and 
matrix models, it
was suggested~\cite{deBoer:2003dn}-\cite{Seiberg:2003yz}, 
that non(anti)commutativity of superspace coordinates 
naturally appears on a $D$-brane worldvolume,
this time due to the
presence of Ramond-Ramond 2-form field strength
in ten dimensions or
equivalently, a self-dual graviphoton field strength 
in four dimensions.
Interestingly, the bosonic coordinates still commute, 
but the fermionic
superspace coordinates do not anticommutate and 
instead, satisfy a
clifford algebra 
$ \{\theta^\alpha,\theta^\beta\} =C^{\alpha \beta}$
~\cite{Casalbuoni:1975bj}-\cite{Alday:2003ms}.
Recently it was shown in~\cite{Seiberg:2003yz},
that such theories with ${\cal N} =1$ supersymmetry 
in $D=4$ are consistent 
and Lorentz invariant at the classical 
level. A surprising aspect is that the 
non(anti)commutativity parameter
$C$ appears explicitly in the superalgebra, and breaks half
of the supersymmetry. Consequently, the 
superspace itself becomes half 
supersymmetric.

The ensuing studies~\cite{Britto:2003aj1}-\cite{Mikulovic:2003sq},
have explored the idea of non(anti)commutativity
of superspace coordinates, in connection with 
matrix models~\cite{Hatsuda:2003ry,Park:2003ku,Shibusa:2003dg}, 
soliton and instanton 
solutions~\cite{Abbaspur:2003ss,Imaanpur:2003jj}, 
UV/IR mixing~\cite{Britto:2003aj} etc..
At the quantum level, sufficient work has been devoted to the 
perturbative
study of supersymmetric gauge theories and Wess-Zumino model with 
${\cal N} =1$ supersymmetry in 
four dimensions~\cite{Terashima:2003ri,Lunin:2003bm,
Berenstein:2003sr,Alishahiha:2003kg}.
Regarding these models,
certain (non)renormalization theorems have been 
proved~\cite{Britto:2003aj} and also
renormalizability~\cite{Grisaru:2003fd,
Britto:2003kg,Romagnoni:2003xt}
to all orders in perturbation theory has been shown.
Furthermore, four dimensional gauge theories with  
$ {\cal N}=2$ supersymmetry 
have also been formulated on a non(anti)commutative 
superspace~\cite{Ivanov:2003te,Ferrara:2003xk}. 

Although, much of the focus has been in connection with the
${\cal N} =1,2$ supersymmetric models in four 
space-time dimensions, it is both
interesting and important to explore the effects of 
non(anti)commutativity
in more general contexts. Especially, $D=2$ offers 
an interesting arena where nonlinear $\sigma$-models 
have been studied 
in great detail. The relation of ultraviolet structure of 
Bosonic $\sigma$-models to the target manifold on which they 
are defined has also been explored~\cite{Friedan:1980jf}.
Recently, 
noncommutative nonlinear $\sigma$-models have also been studied in
considerable detail~\cite{Dabrowski:2000my}-\cite{Murugan:2002rz}.

It is well known that supersymmetric generalization of Bosonic
$\sigma$-models, unveiled  deep
connections between the complex manifold theory and supersymmetry
~\cite{Zumino:1979et}-\cite{Braaten:is}. The
restriction on the type of manifolds compatible with extended supersymmetry
has compelled the study of K\"{a}hler manifolds, besides leading to
the construction of new, previously unknown manifolds~\cite{Lindstrom:rt}. 
To be precise, in connection with the
supersymmetric extension of  Bosonic $\sigma$-models,  it was shown by 
Zumino~\cite{Zumino:1979et}, 
and by Alvarez-Gaum\'{e} and 
Freedman~\cite{Alvarez-Gaume:1980vs}
-\cite{Alvarez-Gaume:hn}, that ${\cal N} =1$ 
supersymmetry is possible on any arbitrary manifold, but $ {\cal N}=2$ 
supersymmetry requires the manifold to be K\"{a}hler and ${\cal N}=4$ 
supersymmetry can only be realized if the manifold is hyperK\"{a}hler.
Remarkably, K\"{a}hler geometry has proved to
have 
strong implications for the renormalizability of $ {\cal N}=2$ supersymmetric
$\sigma$-models defined on either Riemannian or  K\"{a}hler manifolds
~\cite{Alvarez-Gaume:1980dk}-\cite{Alvarez-Gaume:hn}.

Also, the importance of $ {\cal N}=2$ supersymmetric sigma model in the
context of $ {\cal N}=2$ strings is well 
known~\cite{Ademollo:1976pp}-\cite{Ooguri:1991ie}. 
In the recent past,
noncommutative $ {\cal N}=2$ strings have also been
discussed~\cite{Kumar:2000qg,Lechtenfeld:2000nm}.
Motivated by the above
results, in this paper, we explore the implications of 
 K\"{a}hler structure for two dimensional $ {\cal N}=2$
supersymmetric theories on a
non(anti)commutative superspace.

The rest of the paper is organized as follows. In section-2,
we present our  $ {\cal N}=2$ ($D=2$) non(anti)commutative superspace,
while also reviewing certain general properties of  
K\"{a}hler manifolds. 
In section-3, we obtain the classical action for
an $ {\cal N}=2$ supersymmetric theory defined by an arbitrary  
K\"{a}hler potential of a single chiral and antichiral superfield and 
show the emergence of a power 
series expansion in $(\det\,C)$. In section-4, we discuss
the supersymmetry properties of our theory and generalize
the results to include the superpotential. Conclusions and discussion 
are presented in section-5.

\section{\bf{Non(anti)commutative Superspace }}
We start by formulating our 
$ {\cal N}=2$ supersymmetric theory, on a 
non(anti)commutative superspace.
In four dimensions, it was noted~\cite{Klemm:2001yu},
that the algebra of superspace coordinates can be made
non(anti)commutative and also associative only in a
Euclidean space-time. Further in~\cite{Seiberg:2003yz}, it was 
elucidated
that starting from an ${\cal N} =1$ supersymmetric theory,
non(anti)commutative algebra can be consistently formulated
on a Euclidean space. Coming to two
dimensions, there is a well known connection between 
$D=4,\,{\cal N} =1$ and $D=2,\, {\cal N}=2$ models, suggesting 
the existence of an associative non(anti)commutative algebra,
as also pointed out in~\cite{Sako:2003jx}. We now define this
superspace in $D=2$ and continue to use Lorentzian signature as
in~\cite{Seiberg:2003yz}.

\subsection{\bf{  $D=2,\,  {\cal N}=2$ superspace   }}
We begin establishing our notations~\cite{Ademollo:1976pp,West}. 
The superspace coordinates are denoted by:
$\tau,\sigma,\theta^{i}_{\alpha}$~\cite{Ademollo:1976pp}.
Here, $\tau,\sigma$ are
space-time labels and $\theta^{i}_{\alpha}$ 
are N Majorana spinors which
are odd elements of a Grassmann algebra i.e., 
$\theta^{i}_{\alpha} \theta^{i}_{\beta} = 
- \theta^{i}_{\beta}\theta^{i}_{\alpha} $~\cite{Ademollo:1976pp}.
For later convenience, we define the light-cone coordinates:
\be \label{bosons}
\xi = \frac{1}{2}(\tau + \sigma), 
\qquad \zeta =  \frac{1}{2}(\tau - \sigma),
\ee
and for spinor coordinates, one can invoke 
the chirality condition using
$\gamma_5$ in two space-time dimensions~\cite{Ademollo:1976pp}:
\be \label{chirality}
\theta^i~=~ \frac{1}{2}(1-\gamma_5)_{\alpha\beta}
\theta^{i}_{\beta}, \qquad
\chi^i~=~\frac{1}{2}(1+\gamma_5)_{\alpha\beta}
\theta^{i}_{\beta},
\ee
with~\cite{Ademollo:1976pp}:
\be
\gamma_5~=~ \pmatrix{ 1 & 0 \cr
              0 & -1 }.
\ee
In what follows, we will be using the complex version of the spinors 
defined below, which can all be taken to be independent
~\cite{Sako:2003jx}:
\be \label{fermions}
\theta = \theta^1 + i \theta^2, \qquad 
\bar{\theta} = \theta^1 - i \theta^2,
\qquad \chi =  \chi^1 + i\chi^2, \qquad 
\bar{\chi} =  \chi^1 - i\chi^2.
\ee
Hence, using the known results for $D=4,\,{\cal N} =2$~\cite{Seiberg:2003yz},
and $D=2,\, {\cal N}=2$~\cite{Sako:2003jx}, 
the non(anti)commutativity between the Grassmannian coordinates 
\eq{fermions}, can be introduced as shown below:
\be \label{deformation}
\{\theta,\theta \}= C^{00}, \qquad   \{\theta,\chi \} = -C^{01},\qquad
\{\chi,\theta \} = -C^{10},\qquad \{\chi,\chi \} =   C^{11},
\ee
with all other anticommutators of 
$\theta,\chi,\bar{\theta},\bar{\chi}$ vanishing. 
Also, 
\bea \label{nonc}
&&[\bar{\theta},\xi] =~ [\bar{\chi},\xi] = 0, \xx
&&[\bar{\theta},\zeta] =~ [\bar{\chi},\zeta]= 0.
\eea
An immediate consequence of  
\eq{deformation} is that, when functions
of non(anti)commutative coordinates i.e., $\theta,\chi$ are multiplied, 
the result has to be Weyl ordered. This
can be systematically implemented, by introducing a star product between
the non(anti)commutative coordinates, as 
defined below~\cite{Seiberg:2003yz}:
\be \label{star1}
f(\theta,\chi)*g(\theta,\chi) = f(\theta,\chi)
\exp\left(
-{C^{00}\over 2 }\overleftarrow{\partial_{\theta} }
\overrightarrow{\partial_{\theta} }
+{C^{01}\over 2 }\overleftarrow{\partial_{\theta} }
\overrightarrow{\partial_{\chi} }
+{C^{10}\over 2 }\overleftarrow{\partial_{\chi} }
\overrightarrow{\partial_{\theta} }
-{C^{11}\over 2 }\overleftarrow{\partial_{\chi} }
\overrightarrow{\partial_{\chi} }\right)g(\theta,\chi).
\ee
\bea \label{star2}
&&~~~~~~~~~= f(\theta,\chi)~
  \Bigl[~ 1- {C^{00}\over 2 }\overleftarrow{\partial_{\theta} } 
\overrightarrow{\partial_{\theta} }
+{C^{01}\over 2 }\overleftarrow{\partial_{\theta} }
\overrightarrow{\partial_{\chi} }
+{C^{10}\over 2 }\overleftarrow{\partial_{\chi} }
\overrightarrow{\partial_{\theta} }
-{C^{11}\over 2 }\overleftarrow{\partial_{\chi} }
\overrightarrow{\partial_{\chi} } \xx
&&~~~~~~~~~~~~~~~~~~~~~~~ -~\left(\frac{1}{16}\det C\right)
~(~     \overleftarrow{\partial_{\theta}} 
      \overleftarrow{\partial_{\chi}}~-~
      \overleftarrow{\partial_{\chi}} 
      \overleftarrow{\partial_{\theta}}        ~ )
(~     \overrightarrow{\partial_{\theta}} 
      \overrightarrow{\partial_{\chi}}~-~
      \overrightarrow{\partial_{\chi}} 
      \overrightarrow{\partial_{\theta}}     ~ )~\Bigr]~g(\theta,\chi).
\eea
Coming to the algebra of bosonic coordinates \eq{bosons}, it has been noted 
that if we assume,
\be \label{8}
[\xi,\xi] = [\zeta,\zeta] = [\xi,\zeta] = 0,\xx 
\ee
then the supersymmetry remains unbroken, but the supercovariant 
derivatives do not act as derivations and hence, defining antichiral 
fields becomes difficult. Alternatively, as pointed out 
in~\cite{Seiberg:2003yz}, to define antichiral 
superfields one works in the chiral coordinate basis defined as:
\be \label{chiralvar}
\xi^{\pm} = \xi \pm \frac{i}{2}\theta\bar{\theta}, \qquad
\zeta^{\pm} = \zeta \pm \frac{i}{2}\chi\bar{\chi}.
\ee
To be consistent with the choice in eqns. \eq{deformation} and \eq{nonc}, 
we impose the following relations:
\bea \label{10}
&&[\xi^-,\xi^-] = [\zeta^-,\zeta^-] = [\xi^-,\zeta^-] = 0,\xx
&&[\xi^-,\theta]=[\xi^-,\bar{\theta}]=[\xi^-,\chi]
=[\xi^-,\bar{\chi}] =0, \xx 
&&[\zeta^-,\theta]=[\zeta^-,\bar{\theta}]=[\zeta^-,\chi]
=[\zeta^-,\bar{\chi}] =0.
\eea
As a consequence, we get the following identities:
\bea \label{11}
&&[\xi,\xi] = 0 = [\zeta,\zeta], \qquad 
[\xi,\zeta] = \frac{1}{4}\,\bar{\theta}\bar{\chi}C^{01}, \xx
&&[\xi,\theta] =  -\frac{i}{2}\bar{\theta}C^{00},\qquad ~
[\zeta,\chi] =  -\frac{i}{2}\bar{\chi}C^{11}, \xx
&&[\xi,\chi] =  \frac{i}{2}\bar{\theta}C^{01},\qquad ~~~
[\zeta,\theta] =  \frac{i}{2}\bar{\chi}C^{10}. 
\eea
Now, we can write down the 
supercovariant derivatives in the chiral basis
as:
\bea \label{derivatives1}
D_1 &&=~~ -\frac{\partial}{\partial\theta} + i \bar{\theta}
\frac{\partial}{\partial\xi^-}, \qquad 
\bar{D}_1 ~~=~~ -\frac{\partial}{\partial\bar{\theta}},  \\
\label{derivatives2}
D_2 &&=~~~~~ \frac{\partial}{\partial\chi } 
- i \bar{\chi}\frac{\partial}{\partial\zeta^-},\qquad
\bar{D}_2 ~~=~~ \frac{\partial}{\partial\bar{\chi}},
\eea
and the supercharges are given as:
\bea \label{supercharges1}
Q_1 &&=~~ -\frac{\partial}{\partial\theta}, \qquad 
\bar{Q}_1 ~~=~~ -\frac{\partial}{\partial\bar{\theta}} - i \theta
\frac{\partial}{\partial\xi^-}, \\
\label{supercharges2}
Q_2 &&= ~~~~~\frac{\partial}{\partial\chi},  \qquad 
\bar{Q}_2 ~~=~~~~\:\, \frac{\partial}{\partial\bar{\chi}} 
+ i \chi\frac{\partial}{\partial\zeta^-}.
\eea
We can now write down the algebra of supercovariant derivatives 
using eqns. \eq{derivatives1} and \eq{derivatives2} as:
\bea \label{dalgebra}
&&\{D_1, D_1  \}=\{D_2, D_2  \}=\{D_1, \bar{D}_2 \} =
\{D_2, \bar{D}_1 \} =0\\
&&\{\bar{D}_1,\bar{D}_1 \}=
\{\bar{D}_2,\bar{D}_2 \} =0\\
&&\{D_1,\bar{D}_1 \}= -i\partial_{\xi^-} , \qquad
\{D_2,\bar{D}_2 \}=-i\partial_{\zeta^-} .
\eea
The above algebra turns out to be same as in the usual superspace.
In addition, all the supercovariant derivatives anticommute with all
the supercharges. Also, using the definitions given in eqns. 
\eq{supercharges1} and \eq{supercharges2} the supercharges 
can be shown to satisfy:
\bea \label{qalgebra1}
&&\{Q_1, Q_1  \}=\{Q_2, Q_2  \}=\{Q_1,\bar{Q}_2 \}=
\{Q_2,\bar{Q}_1 \} = 0, \\
\label{qalgebra2}
&& \{Q_1,\bar{Q}_1 \}= i\partial_{\xi^-},\qquad 
\{Q_2,\bar{Q}_2 \}= i\partial_{\zeta^-} ,\\
\label{qalgebra3}
&&\{\bar{Q}_1,\bar{Q}_1 \}= -C^{00} \frac{\partial^2}
{\partial_{\xi^-}\partial_{\zeta^-}} , 
\qquad \{\bar{Q}_2,\bar{Q}_2 \}= -C^{11} \frac{\partial^2}
{\partial_{\xi^-}\partial_{\zeta^-}} .
\eea
An important aspect of the above algebra,
as first noted in~\cite{Seiberg:2003yz} 
is that, $\bar{Q}$
is no more a symmetry on the non(anti)commutative space, whereas $Q$ still
continues to be the symmetry. Since, half of the supersymetry is broken,
in our case, the unbroken $Q$ supersymmetry can be termed as 
$N=\frac{2}{2}$ supersymmetry. 

Before proceeding, 
we write down certain identities derived from the star product
\eq{star1}
of Grassmannian coordinates given in eqn. \eq{fermions},
for later use:
\bea \label{tt}
\theta*\theta &&=~ \theta \theta + \frac{1}{2} C^{00}, \qquad
\chi*\chi ~=~ \chi \chi + \frac{1}{2} C^{11}, \\
\label{tc}
\theta*\chi &&=~ (\theta\chi) - \frac{1}{2} C^{01}, \qquad
\chi*\theta ~=~ -(\theta\chi) - \frac{1}{2} C^{10}, \\
\label{ttc}
\theta*\Bigl(\theta\chi\Bigr)    
&&=~ - \theta*\Bigl(\chi\theta\Bigr)~~~~~
~=~\frac{1}{2}\Bigl(\,C^{00} \chi + C^{01}\theta\,\Bigr),\\
\label{cct}
\chi*\Bigl(\chi\theta\Bigr) &&=~
- \chi*\Bigl(\theta\chi\Bigr)~~~~~
~=~\frac{1}{2}\Bigl(\,C^{11} \theta + C^{10}\chi\,\Bigr),\\
\label{tctc}
\Bigl(\theta\chi\Bigr)*\Bigl(\theta\chi\Bigr) &&=~ 
- \Bigl(\chi\theta\Bigr)*\Bigl(\theta\chi\Bigr)~
~=~ -\frac{1}{4}\,(\det\,C).
\eea
To construct a superspace action, we need the chiral and antichiral 
superfields,
which we define below. Chiral superfields satisfy
$\bar{D}_1 S =0=\bar{D}_2 S $ and can be written as:
\be \label{chiral}
S(\xi^-,\zeta^-,\theta,\chi) = A(\xi^-,\zeta^-) 
+ i\theta\bar{\psi}_L(\xi^-,\zeta^-) 
- i\chi\bar{\psi}_R(\xi^-,\zeta^-) - \frac{1}{2}
i\Bigl(\theta*\chi-\chi*\theta\Bigr) F(\xi^-,\zeta^-),
\ee
where the star product of $\theta$ and $\chi$ appearing in 
eqn. \eq{chiral} is already Weyl ordered as shown below:
\be 
\Bigl(\theta*\chi\,-\,\chi*\theta\Bigr) 
= \Bigl(\theta\chi\,-\,\chi\theta\Bigr) = 2 (\theta\chi).
\ee
Now, antichiral superfields satisfying
$D_1 \bar{S} =0=D_2 \bar{S} $ can be written as:
\be \label{antichiral}
\bar{S}(\xi^+,\zeta^+,\bar{\theta},\bar{\chi}) = 
\bar{A}(\xi^+,\zeta^+) 
- i\bar{\theta}\psi_L(\xi^+,\zeta^+) 
+ i\bar{\chi}\psi_R(\xi^+,\zeta^+) 
- i\bar{\theta}\bar{\chi} \bar{F}(\xi^+,\zeta^+).
\ee
It has been pointed out in~\cite{Seiberg:2003yz} that, 
since $(\xi^+,\zeta^+)$ do not commute
among themselves, one needs to Weyl order functions of these coordinates
as well, which is inconvenient. Hence, using the definitions given in 
\eq{chiralvar}, we expand the antichiral 
superfield around  $(\xi^-,\zeta^-)$
coordinates, so that, we only Weyl order the $\theta$'s.
Doing so, one ends up with~\cite{Seiberg:2003yz}:
\bea \label{Sbar}
\bar{S}(\xi^+,\zeta^+,\bar{\theta},\bar{\chi})
&=& \bar{A}(\xi^-,\zeta^-) - i\bar{\theta}\psi_L(\xi^-,\zeta^-) 
+ i\bar{\chi}\psi_R(\xi^-,\zeta^-) + i \theta\bar{\theta}
\partial_{\xi^-}\bar{A} + i\chi\bar{\chi} 
\partial_{\zeta^-}\bar{A} \xx
&+&\bar{\theta}\bar{\chi}\left(-\chi\partial_{\zeta^-}\psi_L
-\theta\partial_{\xi^-}\psi_R -i\bar{F}+(\theta\chi)\partial_{\xi^-}
\partial_{\zeta^-} \bar{A}  \right). 
\eea
Further, the star product does not break chirality. The star product
of two chiral superfields is still chiral, as seen below:
\be \label{2S}
S(\xi^-,\zeta^-,\theta,\chi) * S(\xi^-,\zeta^-,\theta,\chi)
= S(\xi^-,\zeta^-,\theta,\chi)S(\xi^-,\zeta^-,\theta,\chi)
+ \left(\frac{1}{4}\det~C\right) F^2.
\ee
Besides, it is interesting to note the appearance of the 
non(anti)commutativity parameter as $(\det\,C)$. One
can also check that the star product does not 
break antichirality as well, i.e., the
star product of two antichiral 
superfields still remains antichiral.

\subsection{\bf{ $\sigma$-models and K\"{a}hler geometry}}

Here, we give a brief review of the known results concerning
the supersymmetric extension of the Bosonic nonlinear 
$\sigma$-models and also write identities 
coming from the structure of
the K\"{a}hler geometry~\cite{Kobayashi:1963,Ko:gw}.

Let us start by writing down
the action for the bosonic nonlinear $\sigma$-model on a general
even dimensional Riemannian manifold $M$, with a real metric 
$g_{i\bar{j}}(z,\bar{z})$ and the complex coordinate fields 
$z^i,\bar{z}^j$ , where $i,j=1,...,n$, as shown below: 
\be \label{Bsigma}
I = \int d^2x~g_{i\bar{j}} \,
\partial_{\mu}z^{i}\partial_{\mu}\bar{z}^{j}.
\ee
Constraining the manifold to be hermitian, 
the unmixed components of the metric vanish, i.e.,
$g_{ij}= g_{\bar{i}\bar{j}} = 0$ . Moreover, 
as has been already said, 
to couple the  bosonic  $\sigma$-model \eq{Bsigma} to  $n$
complex spinor fields and have $ {\cal N}=2$ supersymmetric extension, the 
target manifold has to be K\"{a}hler. 

The line element on a K\"{a}hler manifold can be written locally as:
\be \label{line}
ds^2 = 2 g_{i\bar{j}}\, dz^{i} d\bar{z}^{j},
\ee
where the hermitian metric $g_{i\bar{j}}$ can be obtained (locally) 
as a second derivative (once holomorphic
and once anti-holomorphic) of an arbitrary
real scalar function (K\"{a}hler potential), 
say, ${\mathcal K}(z,\bar{z})$ as shown:
\be \label{metric}
g_{i\bar{j}} = \frac{\partial}{\partial z^{i}} 
\frac{\partial}{\partial \bar{z}^{j}} {\mathcal K}(z,\bar{z}).
\ee
The above definition of metric \eq{metric}, is invariant 
under  K\"{a}hler gauge transformations and arbitrary 
holomorphic coordinate transformations:
\bea 
&&{\mathcal K}'(z,\bar{z}) \,=\, {\mathcal K}(z,\bar{z}) \,
+\, \Lambda(z)\,+\,\bar{\Lambda}(\bar{z}),  \xx
&&(z^i)' \,=\, z^{'i}(z^j)\:, \qquad
(\bar{z}^i)' \,=\, \bar{z}^{'i}(\bar{z}^j).
\eea
Evidently the metric \eq{metric} is real and off diagonal 
with the only nonzero 
components being $g_{i\bar{j}}$ and $g_{\bar{j}i}$. The non vanishing 
components of the K\"{a}hler metric and their inverses, further satisfy:
\be \label{metinv}
g_{i\bar{j}}\,g^{k\bar{j}} = \delta^{k}_{i}, \qquad
g^{i\bar{j}}g_{i\bar{k}} = \delta^{\bar{j}}_{\bar{k}}.
\ee
From the K\"{a}hler condition, 
\be \label{kahler}
\partial_k g_{i\bar{j}}(z,\bar{z}) = 
\partial_i g_{k\bar{j}}(z,\bar{z}),
\ee
many simplifications occur and very few components of 
the Christoffel symbols and the Riemann Curvature tensor remain nonzero. 
We first write down the non vanishing 
components of 
Christoffel symbols and the 
Riemann Curvature tensor in terms of the  K\"{a}hler potential
${\mathcal K}(z,\bar{z})$:
\bea \label{christo}
&&\Gamma^{i}_{jk} = g^{i\bar{l}} \, \frac{\partial}{\partial z^{j}} 
\frac{\partial}{\partial z^{k}}
\frac{\partial}{\partial \bar{z}^l}\, {\mathcal K}(z,\bar{z}),
\qquad
\Gamma^{\bar i}_{\bar{j}\bar{k}} = g^{\bar{i}l} \, 
\frac{\partial}{\partial \bar{z}^j} 
\frac{\partial}{\partial \bar{z}^k }
\frac{\partial}{\partial z^{l}} \,{\mathcal K}(z,\bar{z}),\xx
&&R^{i}_{j\bar{k}l} =  g^{i\bar{m}} \, 
\frac{\partial}{\partial z^{j}} 
\frac{\partial}{\partial z^{l}}
\frac{\partial}{\partial \bar{z}^k }
\frac{\partial}{\partial \bar{z}^m } \,
{\mathcal K}(z,\bar{z}).
\eea
Starting from the basic definition given in eqn.\eq{metric}, one
can also deduce the following identities:
\bea \label{curv}
&&\Gamma^{i}_{jk} = g^{i\bar{l}}\,\partial_{j}g_{k\bar{l}},\qquad
\Gamma^{\bar{i}}_{\bar{j}\bar{k}} = 
g^{\bar{i}l}\,\partial_{\bar{j}}g_{\bar{k}l},\qquad
\Gamma_{\bar{i}jk} =  g_{\bar{i}l}\,\Gamma^{l}_{jk}, \qquad
\Gamma_{i\bar{j}\bar{k}} =  g_{i\bar{l}}\,
\Gamma^{\bar{l}}_{\bar{j}\bar{k}}, \xx
&&R^{i}_{j\bar{k}l} = \partial_{\bar{k}}\Gamma^{i}_{jl},\qquad
R_{\bar{i}j\bar{k}l} = g_{\bar{i}m}R^m_{j\bar{k}l},
\qquad R_{\bar{i}j\bar{k}l} = R_{\bar{i}l\bar{k}j}.
\eea
Furthermore, the Ricci tensor 
$R_{i\bar{j}}= g^{m\bar{n}}\,R_{\bar{n}i\bar{j}m}$, is a
 ``K\"{a}hler tensor'' as it satisfies $R_{ij}=R_{\bar{i}\bar{j}}=0$
and locally it can be obtained  as
$R_{i\bar{j}} = \partial_{i}\partial_{\bar{j}}\ln \det(g^{k\bar{l}})$.

\section{\bf{$ {\cal N}=2$ Supersymmetric theory on  
Non(anti)commutative Superspace  }}           
In this section, we derive the classical action for 
$ {\cal N}=2$ supersymmetric theories with K\"{a}hler structure on 
non(anti)commutative superspace and then in section-4, 
show that half of the
supersymmetry remains unbroken.
In this paper, we restrict ourselves to a single
chiral and antichiral supermultiplet.
The generalization to include $n$ chiral and 
antichiral superfields remains a subject of future work.

\subsection{\bf{ Classical action  }}           

Continuing the discussion in section-2.2, 
the most general action for 
$ {\cal N}=2$ supersymmetric theories characterized by a 
K\"{a}hler potential ${\mathcal K}(S,\bar{S})$, where
$S$ and $\bar{S}$ are 
chiral and antichiral superfields, 
takes an extremely simple form as shown below:
\be \label{action}
I = \int d^2x \:d\theta \:d\chi\: d\bar{\theta}\: d\bar{\chi}\:
{\mathcal K}(S,\bar{S}),
\ee
where $S$ and $\bar{S}$ are chiral and antichiral superfields
defined in eqns. \eq{chiral} and \eq{Sbar}.
To obtain the action in terms of the component fields, one 
expands ${\mathcal K}(S,\bar{S})$ around the bosonic fields, 
$A$ and $\bar{A}$. Following this general procedure in our case and
using the
definitions of chiral and antichiral superfields given in eqns. \eq{chiral}
and \eq{Sbar}, we first
explicitly write down the possible terms in the expansion of
the  K\"{a}hler potential 
${\mathcal K}(S,\bar{S})$: 
\bea \label{expand}
{\mathcal K}(S,\bar{S}) \!\!\!\!\!\!\! 
&&= {\mathcal K}(A,\bar{A}) \,+\, L \,\frac{\partial 
{\mathcal K} }{\partial S }\,+\,R \,
\frac{\partial{\mathcal K}}
{\partial\bar{S}}
\,+ \frac{1}{2!}\,L*L\,
\frac{\partial^2{\mathcal K}}{\partial S^2}
\,+ \frac{1}{2!}\,R*R\,
\frac{\partial^2{\mathcal K}}{\partial \bar{S}^2} \xx
&&~~\,+ \frac{1}{2!}
\,\left[\,L*R \,\right]\, \frac{\partial^2{\mathcal K}}
{\partial S\partial\bar{S}} 
\,+\,\frac{1}{3!}
\,\left[\,L_*^2*R\,\right] \, 
\frac{\partial^3{\mathcal K}}{\partial S^2\partial\bar{S}} 
\,+\,  \frac{1}{3!}
\, \left[\,R_*^2*L\,\right] \,  
\frac{\partial^3{\mathcal K}}
{\partial S\partial \bar{S}^2} \xx
&&~~+\:\cdots\:+\, 
\frac{1}{n!}\:L_*^{n}\:
\frac{\partial^{n}{\mathcal K}}
{\partial S^{n}}\:+\:\cdots\,+\,
\frac{1}{n!}\:R_*^{n}\:
\frac{\partial^{n}{\mathcal K}}
{\partial \bar{S}^{n}}\:+\:\cdots\, \xx
&&~~
+\,\frac{1}{(n+m)!}
\:\left[\,L_*^{n}*R_*^{m}\,\right] \: 
\frac{\partial^{n+m}{\mathcal K}}
{\partial S^{n}\partial \bar{S^{m} }}\:+\:\cdots\,,
\eea
where $n$ and $m$ are integers and we 
have also introduced the following notations:
\bea \label{LRnota}
&&L = S - A, \qquad R = \bar{S} -\bar{A}, \xx
&&L_*^{n}~=~ \stackrel{n}{\overbrace{L*L*......*L}}, \qquad
R_*^{m}~=~ \stackrel{m}{\overbrace{R*R*......*R}},
\eea
and the square brackets $[\cdots]$ in eqn. \eq{expand}, 
signify 
all possible combinations of $L'$s and $R'$s. For instance,
in this notation: 
$\left[L*L*R\right] \:\equiv \:L*L*R+L*R*L+R*L*L$.

Also, to keep the notations simple in eqn. \eq{expand},
\be
\frac{\partial{\mathcal K} }{\partial S }\, ,\:
\frac{\partial^2{\mathcal K}}{\partial S^2}\, ,\:\cdots\,
\frac{\partial^{n+m}{\mathcal K}}
{\partial S^{n}\partial \bar{S^{m}}}~~{\rm etc.,}\, 
\ee
stand for the derivatives of the K\"{a}hler potential, evaluated
at $S\,=\,A$ and $\bar{S}\,=\,\bar{A}$. 

For $L$ and $R$, explicitly one has:
\bea \label{L}
L(\xi^-,\zeta^-,\theta,\chi)
&& =~ + i\theta\,\bar{\psi}_L - i\chi\,\bar{\psi}_R 
   - i(\theta\,\chi)\, F, \\
\label{R}
R(\xi^+,\zeta^+,\bar{\theta},\bar{\chi})
&& =~- i\bar{\theta}\,\psi_L 
 + i\bar{\chi}\psi_R + i \theta\bar{\theta}
\partial_{\xi^-}\bar{A} + i\chi\bar{\chi} 
\partial_{\zeta^-}\bar{A} \xx
&&~~~+~\bar{\theta}\,\bar{\chi}\left(-\chi\,\partial_{\zeta^-}\psi_L
-\theta\,\partial_{\xi^-}\psi_R -i\bar{F}+
(\theta\,\chi)\,\partial_{\xi^-}
\partial_{\zeta^-} \bar{A}  \right),
\eea
where we have suppressed the dependence of  
component fields on the coordinates  $(\xi^-,\zeta^-)$. 
All the component fields are taken to be functions
of $(\xi^-,\zeta^-)$, if not explicitly mentioned.
Using the form of $L$ and $R$ given in eqns. \eq{L} and \eq{R}, 
we arrive at the following identities:
\bea \label{Lstar}
L_*^{2n}&&=~ \left(\,\frac{1}{4} \det~C\right)^{n-1} \,F^{2n-2}
\,\left[-2n \,(\theta \chi)\:
\bar{\psi}_L\,\bar{\psi}_R \,+\,  
\left(\frac{1}{4}\det~C\right)\, F^2\, \right], \\
\label{L1star}
L_*^{2n+1}&&=~\left(\,\frac{1}{4} \det~C\right)^{n}\, F^{2n-1}\,
\left(-2n\, i\,\bar{\psi}_L\,\bar{\psi}_R \,+\,  F.L\,\right),\\
\label{R2star}
R_*^2 &&=~ 2 \bar{\theta}\,\bar{\chi}\,
\Bigl( -\psi_L\psi_R \,-\,\chi\,\psi_L
  \,\partial_{\zeta^-} \bar{A}\,
  -\,\theta\,\psi_R\,\partial_{\xi^-} \bar{A}\,         
+\,(\theta\,\chi)\,\partial_{\xi^-} \bar{A}\partial_{\zeta^-} \bar{A}
\Bigr),  \\
\label{Rnstar}
R_*^{n}&&=~ 0,\quad {\rm for}~~ n \,>\, 2.
\eea

To evaluate the action in eqn. \eq{action} explicitly, one 
needs to collect the terms with coefficient 
$\,\bar{\theta}\,\bar{\chi}\,(\theta\,\chi)\,$ appearing in 
the expansion of K\"{a}hler potential in eqn. \eq{expand}, as 
only these terms survive after 
performing integration over the Grassmann variables.
Now, from eqns. \eq{L}, \eq{Lstar} and \eq{L1star}, 
one can see that the star
product of an arbitrary number of $L'$s alone does not have a term 
proportional to $\,\bar{\theta}\,\bar{\chi}\,(\theta\,\chi)\,$,
and hence, it cannot give rise
to terms in the action. On the other hand, 
with the same logic, $R$ given in eqn. \eq{R}
the star product of two  $R'$s given in eqn. \eq{R2star},
do contribute terms to the action. Further, 
the star product of three and higher 
number of $R'$s vanishes and hence, does 
not contribute to the action.
Therefore, to proceed, we
turn towards the star product of an arbitrary number of $L'$s
with either  $R$ or $R*R$. 
First, using  the definitions of $L$ and $R$ given in 
eqns. \eq{L} and \eq{R}, we arrive at the following relation:
\bea \label{LRstar}
L\,*\,R \,+\, R\,*\,L &&=
~ -\,2\,\bar{\chi}\,\theta\,\bar{\psi}_L\,\psi_R \,-\,2\,
\bar{\theta}\,\chi\,\bar{\psi}_R\,\psi_L \,+\,2\,
\bar{\theta}\,\theta\,\bar{\psi}_L\,\psi_L \,+\,2\,
\bar{\chi}\,\chi\,\bar{\psi}_R\,\psi_R \, \xx
&&~~~~+\,\frac{i}{2}\:
\bar{\theta}\,\bar{\chi}\,\left(\det~C \right)\,F\,
\partial_{\xi^-}\partial_{\zeta^-}\bar{A} \,-\,2\,
\bar{\theta}\,(\theta\,\chi)\,\left(\bar{\psi}_R\,
\partial_{\xi^-}\bar{A} \,+\,F\,\psi_L \,\right)\, \xx 
&&~~~~-\,2\,\bar{\chi}\,(\theta\,\chi)\,
\left(\bar{\psi}_L\,
\partial_{\zeta^-}\bar{A} \,-\,F\,\psi_R\right)\,
+\,2\,\bar{\theta}\,\bar{\chi}\,\theta\,\bar{\psi}_L\,\bar{F}\,-\,
2\,\bar{\theta}\,\bar{\chi}\,\chi\,\bar{\psi}_R\,\bar{F}\,\xx
&&~~~~~~+\,2\,\bar{\theta}\,\bar{\chi}\,(\theta\,\chi)\,
\left( i\,\bar{\psi}_L\,\partial_{\zeta^-}\psi_L
\,+\, i\,\bar{\psi}_R\,
\partial_{\xi^-}\psi_R \,-\, F\,{\bar F}\right).
\eea
Note that in the above equation, $L*R$ and $R*L$ independently
have many more terms (explicit expressions are given
in the Appendix)
which cancel out when we consider 
$\Bigl[\,L\,*\,R \,\Bigr]$. Further, since our 
interest is in extracting only the terms with coefficient
$\,\bar{\theta}\,\bar{\chi}\,(\theta\,\chi)\,$, we mainly
concentrate on those terms. Moreover, in this connection, we 
make the observation that, focusing 
only on the terms with coefficient 
$\,\bar{\theta}\,\bar{\chi}\,(\theta\,\chi)\,$, a great deal
of simplification occurs, allowing us to obtain
several identities which we give in the Appendix.
The  crux of the matter is that a general term
in the expansion in eqn. (\ref{expand}), e.g., 
$\left[ L_*^{p}*R* 
L_*^{q}*R* L_*^{r}\right]$, can give rise 
to various terms in the action 
corresponding to all possible 
combinations of $L$'s and $R$'s.
However, using the identities given in the Appendix,
one can push all the $L$'s to one side and all
the $R$'s to the other side as:
\be \label{combi}
\left[ L_*^{p}*R* L_*^{q}
*R*L_*^{r}\right]|_{\bar{\theta}
\,\bar{\chi}\,\theta\,\chi\,} 
~\equiv ~ \:{^{(p+q+r+2)}C_2}\: 
L_*^{(p+q+r)}*R_*^2\,|_{\bar{\theta}
\,\bar{\chi}\,\theta\,\chi\,} , 
\ee
where $p,q,r$ are integers.
We give the proof of these identities  
in the Appendix and in what follows use these results
directly. 
Explicit check of supersymmetry (in section-4) further 
confirms that our action is indeed correct. 

The star product of even and odd number of
$L'$s with  
$R$, can be shown to satisfy (when restricted
to terms with 
coefficient $\bar{\theta}\,\bar{\chi}\,(\theta\,\chi)\,)$:
\bea
\label{LnstaR}
\left[\,L_*^{2n}\,*\,R\,\right]|_{\bar{\theta}
\,\bar{\chi}\,\theta\,\chi\,} 
&&=~~\:{^{(2n+1)}C_{1}}\,\,L_*^{2n}\,*\,R\,|_{\bar{\theta}
\,\bar{\chi}\,\theta\,\chi\,}  \xx
&&=~~
 \:(2n+1)\,
\left(\frac{1}{4}\det~C\right)^{n-1}\,F^{2n-2}
\Bigl[\, 2n\,i \,
\bar{\psi}_L\,\bar{\psi}_R\,\bar{F} \, \xx
&&~~~~~+\, \,  
\left(\frac{1}{4}\det~C\right)\,F^{2}\,
\partial_{\xi^-}\partial_{\zeta^-}\bar{A} \,\Bigr],\\
\label{Ln1staR}
\left[\,L_*^{2n+1}\,*\,R\,\right]|_{\bar{\theta}\,
\bar{\chi}\,\theta\,\chi\,}
&&=~~\:{^{(2n+2)}C_{1}}\,\,L_*^{2n+1}\,*\,R\,|_{\bar{\theta}
\,\bar{\chi}\,\theta\,\chi\,}  \xx
&&=~~\,
 \:(2n+2)\,
\left(\frac{1}{4}\det~C\right)^{n}\,F^{2n-1}\,
\Bigl[\,-2n\,i\,\bar{\psi}_L\,\bar{\psi}_R\, 
\partial_{\xi^-}\partial_{\zeta^-}\bar{A}  \xx
&&~~~~~+\, \,F\,
\left( \,i\,\bar{\psi}_L\,\partial_{\zeta^-}\psi_L\,
\,+\, i\,\bar{\psi}_R\,
\partial_{\xi^-}\psi_R \,-\, F\,{\bar F}\,
\right)\, \Bigr],
\eea
where $n \,\geq 1$. Now,
one can notice that  eqn. \eq{LRstar}, has terms proportional to
$\,\bar{\theta}\,\bar{\chi}\,(\theta\,\chi)\,$ 
which are also independent of $\,C$. Therefore, these terms
would give rise to the standard $C=0$, 
$ {\cal N}=2$ supersymmetric action. On the other hand,
from the set of eqns. \eq{LnstaR} and \eq{Ln1staR}, and similar
ones appearing below,
one can see the emergence of new
terms in the action, 
proportional to arbitrary powers of $(\det\,C)$.

Next, we go on to compute the star product of arbitrary powers of 
$L'$s with $R_*^2$, needed for writing down our action. 
Using the results given for $L'$s in
eqns. \eq{Lstar} and \eq{L1star} and $R_*^2$ in \eq{R2star},
we end up with:
\bea \label{LstaRR}
\left[\,L*R_*^2\,\right]|_{\bar{\theta}\,
\bar{\chi}\,\theta\,\chi\,}
&&=~~\:{^{3}C_{1}}\,\,L*R_*^2\,|_{\bar{\theta}
\,\bar{\chi}\,\theta\,\chi\,}  \xx
&&=~~\,6\,i\,
\left(\psi_L\,\psi_R \,F \,
+\,\bar{\psi}_L\,\psi_L\,\partial_{\zeta^-}\bar{A}\,
+\,\bar{\psi}_R\,\psi_R\,\partial_{\xi^-}\bar{A}\,
\right), \\
\label{LnstaRR}
\left[\,L_*^{2n}*R_*^2\,\right]|_{\bar{\theta}\,
\bar{\chi}\,\theta\,\chi\,}
&&=~~ \:{^{(2n+2)}C_{2}}\,\,L_*^{2n}*R_*^2\,|_{\bar{\theta}
\,\bar{\chi}\,\theta\,\chi\,}  \xx
&&=~~\, 
 \:{^{(2n+2)}C_{2}}\,2\,
\left(\frac{1}{4}\det~C\right)^{n-1}\,F^{2n-2}\,\Bigl[\,2\,n
\bar{\psi}_L\,\bar{\psi}_R\,\psi_L\,\psi_R \,\xx
&&~~~~~~+\,  
\left(\frac{1}{4}\det~C\right)\,F^{2}\,
\partial_{\xi^-}\bar{A}\,\partial_{\zeta^-}\bar{A}\,\Bigr], \\
\label{Ln1staRR}
\left[\,L_*^{2n+1}*R_*^2\,\right]|_{\bar{\theta}\,
\bar{\chi}\,\theta\,\chi\,}
&&=~~\:{^{(2n+3)}C_{2}}\,\,L_*^{2n+1}*R_*^2 |_{\bar{\theta}
\,\bar{\chi}\,\theta\,\chi\,} \xx
&&=~~ 
-\,
 \:{^{(2n+3)}C_{2}}\,2\,i\,
\left(\frac{1}{4}\det~C\right)^{n}\,F^{2n-1}\,
\Bigl[\,2\,n\,\bar{\psi}_L\,\bar{\psi}_R\, 
\partial_{\xi^-}\bar{A}\,\partial_{\zeta^-}\bar{A} \xx
&&~~~~~-\,F\,\left(  \,\bar{\psi}_L\,\psi_L\,
\partial_{\zeta^-}\bar{A}\,+\, \bar{\psi}_R\,\psi_R\,
\partial_{\xi^-}\bar{A}\, +\,\psi_L\,\psi_R\, F\,\right)\, \Bigr],
\eea
where $n\,\geq 1$. 
It may be noted that eqn. (\ref{LstaRR}) contributes terms
in the standard $C=0$, supersymmetric
$ {\cal N}=2$ action. 
Furthermore, eqns. \eq{LnstaRR} and \eq{Ln1staRR} 
contribute new terms in the action, proportional to 
various powers of $(\det\,C)$.

To summarize, substituting the results obtained in eqns. 
\eq{L}-\eq{Ln1staRR} in the action
\eq{action}, we end up with the classical action  
for supersymmetric $ {\cal N}=2$ theory
on a non(anti)commuta-\\
tive superspace. It is possible to break the action into 
two parts as:
\be \label{break}
I \,=\, I_0\,+\, I^n_C,
\ee
where $I_0$ corresponds to the  $C=0$,
$ {\cal N}=2$ action 
and $I^n_C$ stands for the non(anti)commutative part of the
$ {\cal N}=2$ theory. 

Below, we first collect the 
terms coming from the expansion of the K\"{a}hler potential
in the action in eqn. \eq{action},
which will ultimately give rise to $I_0$:
\bea
I_0 &&=~ \int d^2x \:d\theta \:d\chi\: 
d\bar{\theta}\: d\bar{\chi}\:
\Bigl[\:
R \,\frac{\partial{\mathcal K}}{\partial\bar{S}}
\,+\,\frac{1}{2!}\,R_*^2\,
\frac{\partial^2{\mathcal K}}{\partial \bar{S}^2}
\,+\, \frac{1}{2!}
\,\left[\,L*R \,\right]\, \frac{\partial^2{\mathcal K}}
{\partial S\partial\bar{S}}             \xx
&&~~~~+\,\frac{1}{3!}
\,\left[\,L_*^2*R\,\right] \, 
\frac{\partial^3{\mathcal K}}{\partial S^2\partial\bar{S}} 
\,+\,  \frac{1}{3!}
\, \left[\,R_*^2*L\,\right] \,  
\frac{\partial^3{\mathcal K}}
{\partial S\partial \bar{S}^2}    
\,+\, \frac{1}{4!}
\, \left[\,L_*^2*R_*^2\,\right] \,  
\frac{\partial^4{\mathcal K}}
{\partial S^2\partial \bar{S}^2}
\:\Bigr].
\eea
Now, making use of the identities derived in 
eqns. (\ref{R}), (\ref{R2star}),
(\ref{LRstar}), (\ref{LnstaR}), (\ref{LstaRR}) and
(\ref{LnstaRR}), in the above form of the action and 
after performing integration 
over the Grassmannian coordinates, in the usual way,
we end up with the following action for $C=0$,
$ {\cal N}=2$ supersymmetric theory:
\bea \label{Ic0p}
&&I_0= \int d^2x \,\Bigl[\,
\Bigl(\,\partial_{\xi^-}\partial_{\zeta^-}\bar{A} 
\,\Bigr)\:
\frac{\partial{\mathcal K}}{\partial \bar{S} }\,+\,
\Bigl(\:i\bar{\psi}_L\partial_{\zeta^-}\psi_L
+ i\bar{\psi}_R\partial_{\xi^-}\psi_R - F{\bar F}  
\:\Bigr)
~\frac{\partial^2{\mathcal K}}{\partial S\partial\bar{S}} \xx
&&~~~~~~+\,
\Bigl(\,
\partial_{\xi^-}\bar{A}\partial_{\zeta^-}\bar{A}
\,\Bigr)\:
\frac{\partial^2{\mathcal K}}{\partial\bar{S}^2}
+ i\left(~\bar{\psi}_L\bar{\psi}_R{\bar F}~\right)~ 
\frac{\partial^3{\mathcal K}} {\partial S^2\partial\bar{S}} 
+ i\left(~\psi_L\psi_R F 
+\bar{\psi}_L\psi_L\partial_{\zeta^-}\bar{A} \right.\xx
&&~~~~~~+ \left. \bar{\psi}_R\psi_R\partial_{\xi^-}\bar{A}~\right)
~\frac{\partial^3{\mathcal K}}{\partial S\partial \bar{S}^2} \,
-\, \left(~\bar{\psi}_L\psi_L \bar{\psi}_R\psi_R  ~\right)~
\frac{\partial^4{\mathcal K}}{\partial S^2\partial \bar{S}^2}\,\Bigr].
\eea
Now, to write the action in the standard form, we perform
a partial integration 
on the $\frac{\partial {\mathcal K}}{\partial S}$ term. The
action so obtained is given below:
\bea \label{Iczero}
&&I'_0= \int d^2x \,\Bigl[\,
\Bigl(~-\frac{1}{2}\partial_{\xi^-}A\partial_{\zeta^-}\bar{A}
-\frac{1}{2}\partial_{\zeta^-}A\partial_{\xi^-}\bar{A}
+i\bar{\psi}_L\partial_{\zeta^-}\psi_L
+ i\bar{\psi}_R\partial_{\xi^-}\psi_R - F{\bar F}  ~\Bigr)
~\frac{\partial^2{\mathcal K}}{\partial S\partial\bar{S}} \xx
&&~~~~~~+ i\left(~\bar{\psi}_L\bar{\psi}_R{\bar F}~\right)~ 
\frac{\partial^3{\mathcal K}} {\partial S^2\partial\bar{S}} 
+ i\left(~\psi_L\psi_R F +\bar{\psi}_L\psi_L\partial_{\zeta^-}\bar{A}
+  \bar{\psi}_R\psi_R\partial_{\xi^-}\bar{A}~\right)
~\frac{\partial^3{\mathcal K}}{\partial S\partial \bar{S}^2} \xx
&&~~~~~~- \left(~\bar{\psi}_L\psi_L \bar{\psi}_R\psi_R  ~\right)~
\frac{\partial^4{\mathcal K}}{\partial S^2\partial \bar{S}^2}\,\Bigr].
\eea 
This should be compared with the usual $C=0,\, {\cal N}=2$ 
supersymmetric action~\cite{Braaten:is}, which can be determined
by taking the superspace coordinates to anticommutate.

Before proceeding to write down the action, we first
define the chiral and antichiral superfields in the $C=0$ 
{\em anticommutative} theory:
\bea \label{Phi}
\tilde{S}(\xi^-,\zeta^-,\theta,\chi) 
&& =~ \tilde{A}(\xi,\zeta) 
+ i\theta\,\bar{\tilde{\psi}}_L(\xi,\zeta) 
 - i\chi\bar{\tilde{\psi}}_R(\xi,\zeta)
 - \frac{i}{2} \theta\bar{\theta}
\partial_{\xi} \tilde{A} 
- \frac{i}{2}\chi\bar{\chi} 
\partial_{\zeta}\tilde{A}              \xx
&&~~~+~\theta\,\chi\left(\frac{\bar{\chi}}{2}\,
\partial_{\zeta}\bar{\tilde{\psi}}_L
+\frac{\bar{\theta}}{2}\,
\partial_{\xi}\bar{\tilde{\psi}}_R -i\,\tilde{F}+
\frac{1}{4}\bar{\theta}\,\bar{\chi}\,\partial_{\xi}
\partial_{\zeta} \tilde{A}  \right)    \\ %
\label{Phibar}
\bar{\tilde{S}}(\xi^+,\zeta^+,\bar{\theta},\bar{\chi})
&& =~\bar{\tilde{A}}(\xi,\zeta)- i\bar{\theta}\,\tilde{\psi}_L 
 + i\bar{\chi}\tilde{\psi}_R + \frac{i}{2} \theta\bar{\theta}
\partial_{\xi}\bar{\tilde{A}} + \frac{i}{2}\chi\bar{\chi} 
\partial_{\zeta}\bar{\tilde{A}}           \xx
&&~~~+~\bar{\theta}\,\bar{\chi}
\left(-\frac{\chi}{2}\,\partial_{\zeta}\tilde{\psi}_L
-\frac{\theta}{2}\,\partial_{\xi}\tilde{\psi}_R 
-i\bar{\tilde{F}}+ \frac{1}{4}
\theta\,\chi\,\partial_{\xi}
\partial_{\zeta} \bar{\tilde{A}}  \right).
\eea
Hence, using the definitions
of chiral and antichiral superfields given in eqns. \eq{Phi} 
and \eq{Phibar},
the classical action can be constructed 
by expanding the K\"{a}hler potential of the 
{\em anticommutative} 
theory. Below, we only give the
final result for the 
action, after performing partial integrations on the terms 
of the type 
$\frac{\partial {\mathcal K}}{\partial \tilde{S}}$ and
$\frac{\partial {\mathcal K}}{\partial \bar{\tilde{S}} }$:
\bea \label{Iczeroprime}
&&\tilde{I}_0 = \int d^2x \,\left[\, 
\Bigl\{~-\frac{1}{2}\partial_{\xi}\tilde{A}
\partial_{\zeta}\bar{\tilde{A}}
-\frac{1}{2}\partial_{\zeta}\tilde{A}
\partial_{\xi}\bar{\tilde{A}}
+\,\frac{i}{2}
\Bigl(
\bar{\tilde{\psi}}_L\partial_{\zeta}
\tilde{\psi}_L - \partial_{\zeta}
\bar{\tilde{\psi}}_L\tilde{\psi}_L 
\Bigr)
+ \,\frac{i}{2}
\Bigl(
\bar{\tilde{\psi}}_R
\partial_{\xi}\tilde{\psi}_R  \right. \xx
&&~~~~\left.- \partial_{\xi}\bar{\tilde{\psi}}_R
\tilde{\psi}_R 
\Bigr)  
- F{\bar F}  ~\Bigr\}
~\frac{\partial^2{\mathcal K}}
{\partial \tilde{S}\partial\bar{\tilde{S}}} 
+i\,\left(~\bar{\tilde{\psi}}_L
\bar{\tilde{\psi}}_R{\bar F} 
-\frac{1}{2}\bar{\tilde{\psi}}_L
\tilde{\psi}_L\partial_{\zeta}\tilde{A}
- \frac{1}{2} \bar{\tilde{\psi}}_R
\tilde{\psi}_R\partial_{\xi}\tilde{A}
~\right)~ \frac{\partial^3{\mathcal K}} 
{\partial \tilde{S}^2\partial\bar{\tilde{S}}}  \right. \xx
&&~~~~~~~~~~\left.+\: i\,\Bigl(~\tilde{\psi}_L\tilde{\psi}_R F  
+\frac{1}{2}\bar{\tilde{\psi}}_L
\tilde{\psi}_L\partial_{\zeta}\bar{\tilde{A}}
+ \frac{1}{2} \bar{\tilde{\psi}}_R
\tilde{\psi}_R\partial_{\xi}\bar{\tilde{A}}~\Bigr)
~\frac{\partial^3{\mathcal K}}
{\partial \tilde{S}\partial \bar{\tilde{S}}^2} \right. \xx
&&~~~~~~~~~~\left.- \:\left(~\bar{\tilde{\psi}}_L\tilde{\psi}_L 
\bar{\tilde{\psi}}_R\tilde{\psi}_R  ~\right)~
\frac{\partial^4{\mathcal K}}
{\partial \tilde{S}^2\partial \bar{\tilde{S}}^2}\,\right].
\eea
One can notice a slight mismatch in the form of 
$I'_0$ given in \eq{Iczero} and $\tilde{I}_0$
given in \eq{Iczeroprime}.
In particular, the 
$A$,$\bar{A}$ and $\tilde{A}$ ,$\bar 
{\tilde{A}}$  dependent terms with 
coefficient $\frac{\partial^3{\mathcal K}} 
{\partial \tilde{S}^2\partial\bar{\tilde{S}}}$ 
and  $\frac{\partial^3{\mathcal K}} 
{\partial \tilde{S}\partial\bar{\tilde{S}}^2}$ in 
eqn. \eq{Iczeroprime}, are
different in both the cases. However,  the equivalence of the 
actions $I'_0$ and $\tilde{I}_0$, up to a total derivative, can
be explicitly seen by adding the following total derivative
terms to the action in eqn. \eq{Iczeroprime}: 
\be
\frac{i}{2}\partial_{\zeta}\left(\bar{\tilde{\psi}}_L\tilde{\psi}_L
\frac{\partial^2{\mathcal K}} 
{\partial \tilde{S}\partial\bar{\tilde{S}}} \right)  
+\frac{i}{2}\partial_{\xi}\left(\bar{\tilde{\psi}}_R\tilde{\psi}_R
\frac{\partial^2{\mathcal K}} 
{\partial \tilde{S}\partial\bar{\tilde{S}}} \right).  
\ee
The apparent difference in the two actions is only because 
$I'_0$ is written in chiral coordinates $(\xi^-,\zeta^-)$, and
$\tilde{I}_0$ is written in ordinary coordinates  $(\xi,\zeta)$. 

Coming to the non(anti)commutative part of the 
action $I^n_C$, one has to  
follow similar steps as discussed 
above. Using the results derived in 
eqns. \eq{LnstaR}-\eq{Ln1staRR}, various terms 
in the expansion of the K\"{a}hler
potential in the action in \eq{action}, which can
contribute to terms in $I^n_C$ are summarized below:
\bea \label{Icexpand}
I^n_C &&=~ \int d^2x \:d\theta \:d\chi\: 
d\bar{\theta}\: d\bar{\chi}\: \xx
&&~~ \sum_{n=1}^{\infty}
\Bigl[\:
 \frac{1}{(2n+1)!}
\: \left[\,L_*^{2n}*R\,\right]\: 
\frac{\partial^{2n+1}{\mathcal K}}
{\partial S^{2n}\partial \bar{S}} \,
+\, \frac{1}{(2n+2)!}
\:\left[\,L_*^{2n+1}*R\,\right]\: 
\frac{\partial^{2n+2}{\mathcal K}}
{\partial S^{2n+1}\partial \bar{S}} \xx
&&~+\, \frac{1}{(2n+2)!}
\:\left[\,L_*^{2n}*R_*^2\,\right] \: 
\frac{\partial^{2n+2}{\mathcal K}}
{\partial S^{2n}\partial \bar{S^{2} }}
\,+\, \frac{1}{(2n+3)!}
\:\left[\,L_*^{2n+1}*R_*^2\,\right] \: 
\frac{\partial^{2n+3}{\mathcal K}}
{\partial S^{2n+1}\partial \bar{S^{2} }} 
\:\Bigr].
\eea
Now, it is straightforward to substitute the results derived in 
eqns.\eq{LnstaR}-\eq{Ln1staRR}, 
and collect the terms
proportional to various powers of $(\det\,C)$. 
Then,
performing an integration over the Grassmannian variables in the
usual way, one will end up with the final form of action.
In this context we first
collect terms which are to lowest order in $(\det\,C)$ and write
down the action explicitly to first order in $(\det\,C)$. 
The subsequent generalization to
include terms  proportional to arbitrary powers of $(\det\,C)$ 
will follow.

The non(anti)commutative
part of the action
obtained from eqn. \eq{Icexpand} 
to first order in $(\det\,C)$ (labeled as $I_{C1}$) turns out to be:
\bea \label{Icnonzero1}
&&I_{C1} =  \left(\frac{1}{4}\det\,C \right)\int d^2x
\Bigl[
~\left(
~\frac{1}{2} F^2 \partial_{\xi^-} 
\partial_{\zeta^-}\bar{A}
~\right)
\frac{\partial^3{\mathcal K}} 
{\partial S^2\partial\bar{S}}+\frac{F}{6}
\Bigl(~-2i\bar{\psi}_L\bar{\psi}_R\partial_{\xi^-} 
\partial_{\zeta^-}\bar{A} + i
\bar{\psi}_L\partial_{\zeta^-}\psi_L F \xx 
&&~~~~~+ i\bar{\psi}_R
\partial_{\xi^-}\psi_R F - F^2{\bar F}
~\Bigr) 
\frac{\partial^4{\mathcal K}} 
{\partial S^3\partial\bar{S}} 
+\frac{1}{2}\left(~
F^2 \partial_{\xi^-}\bar{A}
\partial_{\zeta^-}\bar{A}
~\right) 
\frac{\partial^4{\mathcal K}} 
{\partial S^2\partial\bar{S}^2} 
+\frac{i}{6} 
\Bigl(~
\bar{\psi}_L\bar{\psi}_R F^2{\bar F} 
~\Bigr) 
\frac{\partial^5{\mathcal K}} 
{\partial S^4\partial\bar{S}} \xx
&&~~~~~
-\frac{i}{6}F~
\Bigl(~2
\bar{\psi}_L\bar{\psi}_R \partial_{\xi^-}
\bar{A}\partial_{\zeta^-}\bar{A} - F^2\psi_L\psi_R 
+ F \psi_L\bar{\psi}_L
 \partial_{\zeta^-}\bar{A} 
+ F \psi_R\bar{\psi}_R \partial_{\xi^-}\bar{A}
~\Bigr)
\frac{\partial^5{\mathcal K}} 
{\partial S^3\partial\bar{S}^2} \xx
&&~~~~~~-\frac{1}{6}
\left(~
F^2\bar{\psi}_L\psi_L\bar{\psi}_R\psi_R
~\right) 
\frac{\partial^6{\mathcal K}} 
{\partial S^4\partial\bar{S}^2} 
~\Bigr].
\eea
As was first pointed in~\cite{Seiberg:2003yz},  $C$ by itself breaks
Lorentz invariance, but since it appears in the action \eq{Icnonzero1},
as $(\det\,C)$ no Lorentz invariance is broken.
Hence, the non(anti)commutative
part of the action \eq{Icnonzero1}, is Lorentz invariant 
to first order in $(\det\,C)$.  

Having written the action to first order in $(\det\,C)$ 
in eqn. \eq{Icnonzero1}, we now proceed to 
write the most general action,
by following similar steps as discussed above. We substitute the 
results obtained in  
eqns. \eq{LnstaR}-\eq{Ln1staRR}, in the action \eq{action}, and 
collect terms which are proportional to  $(\det\,C)^n$.
The most general action for 
non(anti)commutative part of the $ {\cal N}=2$ supersymmetric
theory,
then looks as follows:
\bea \label{Icnonzerofull}
&&I^n_C = \sum_{n= 2}^{\infty}\,\left(\frac{1}{4}
\det\,C \right)^{n-1}\:\int d^2x \:
 F^{2n-2} \,\frac{2n}{(2n)!}
\Bigl[\:i
\bar{\psi}_L\bar{\psi}_R {\bar F} \,
\frac{\partial^{2n+1}{\mathcal K}} 
{\partial S^{2n}\partial\bar{S}} 
\,-\,\bar{\psi}_L\psi_L\bar{\psi}_R\psi_R \,
\frac{\partial^{2n+2}{\mathcal K}} 
{\partial S^{2n}\partial\bar{S}^2} 
\:\Bigr] \xx
&&~~~+  \sum_{n=1}^{\infty}
\left(\frac{1}{4}\det\,C \right)^n  
\:\int d^2x \:  F^{2n-2}
\Bigl[ \,
\frac{1}{(2n)!} F^2 \partial_{\xi^-} 
\partial_{\zeta^-}\bar{A}\,
\frac{\partial^{2n+1}{\mathcal K}} 
{\partial S^{2n}\partial\bar{S}}  \xx
&&~~~-\frac{ F}{(2n+1)!} 
\Bigl(\,2n\,i\bar{\psi}_L\bar{\psi}_R\partial_{\xi^-} 
\partial_{\zeta^-}\bar{A} 
- i\bar{\psi}_L\partial_{\zeta^-}\psi_L F 
- i\bar{\psi}_R\partial_{\xi^-}\psi_R F + F^2{\bar F}
\,\Bigr) 
\frac{\partial^{2n+2}{\mathcal K}} {\partial S^{2n+1}
\partial\bar{S}}  \xx
&&~~~+ \frac{1}{(2n)!}\,
F^{2} \partial_{\xi^-}\bar{A}\partial_{\zeta^-}\bar{A}\,
\frac{\partial^{2n+2}{\mathcal K}} 
{\partial S^{2n}\partial\bar{S}^2} 
-\frac{i}{(2n+1)!}F\,
\Bigl(\,2n
\bar{\psi}_L\bar{\psi}_R \partial_{\xi^-}
\bar{A}\partial_{\zeta^-}\bar{A} - F^2\psi_L\psi_R \xx
&&~~~+ F \psi_L\bar{\psi}_L
 \partial_{\zeta^-}\bar{A} 
+ F \psi_R\bar{\psi}_R \partial_{\xi^-}\bar{A}
\,\Bigr)
\frac{\partial^{2n+3}{\mathcal K}} {\partial S^{2n+1}
\partial\bar{S}^2}
\:\Bigr] \,.
\eea
One can clearly see 
a power series expansion in $(\det\,C)$. The full action 
for the  
$ {\cal N}=2$ supersymmetric theory on a non(anti)commutative superspace 
is thus given by  eqns. \eq{Iczero} and \eq{Icnonzerofull}.
We now proceed to analyze the supersymmetry of the theory.

\section{\bf{Supersymmetry}}           

The two supersymmetries preserved by the usual 
$C=0,\, {\cal N}=2$ supersymmetric theories
are generated by the supercharges $Q$ and  $\bar{Q}$.
However, from the discussion in section-1, we know that, once we 
invoke non(anti)commutativity of the superspace coordinates as in 
eqn. \eq{deformation},
the $\bar{Q}$ symmetry is broken and the superspace becomes 
$N=\frac{2}{2}$
supersymmetric. Accordingly, we shall use the unbroken $Q$ symmetry 
on the superfields to generate the supersymmetry transformations. 

Thus, the variation of the chiral superfield in 
eqn. \eq{chiral} is given as:
\be \label{Svar}
\delta_{\alpha} S \,=\, \delta_{\alpha}A \,+\, i\,\theta\,
\delta_{\alpha} \bar{\psi}_L 
\,-\, i\,\chi\,\delta_{\alpha}\bar{\psi}_R \,-\, i\,
(\theta\,\chi)\,\delta_{\alpha}\, F. 
\ee
As stated before,
we have suppressed the explicit dependence of component fields
on the coordinates for clarity. All the component fields 
are taken to be functions of $(\xi^-,\zeta^-)$ unless specified
otherwise. The  supercharge $Q$ acts on the 
chiral superfield as shown:
\bea  \label{Sqvar}
\delta_{\alpha} S &=& \left(~\alpha_1\, Q_1 \,+\, 
\alpha_2\, Q_2~\right) \,*\, S, \xx
&=& -\alpha_1 \,\left(\, i\,\bar{\psi}_L \,-\, i\,
\chi\, F \,\right) \,+\, \alpha_2\, \left(\, 
\,-\,i\,\bar{\psi}_R \,+\, i\,\theta\, F \,\right).
\eea
Now, it is easy to compare eqns. \eq{Svar} and \eq{Sqvar}, 
and get the supersymmetry transformations
for the component fields: 
\bea \label{susy}
\delta_{\alpha}A &=& -i\,\alpha_1 \,\bar{\psi}_L 
\,-\, i\,\alpha_2 \,\bar{\psi}_R, \xx 
\delta_{\alpha}\bar{\psi}_L &=& -\alpha_2\, F,  \xx 
\delta_{\alpha}\bar{\psi}_R &=& \alpha_1\, F, \xx
\delta_{\alpha} F &=& 0.
\eea
Analogously, the supersymmetry transformations of the remaining
component fields appearing in the antichiral superfield in 
eqn. \eq{Sbar} can also be written:
\bea \label{susybar}
\delta_{\alpha}\bar{A} &=& 0,\xx
\delta_{\alpha}\psi_L &=& 
-\alpha_1\,\partial_{\xi^-}\bar{A}, \xx
\delta_{\alpha}\psi_R &=& 
-\alpha_2 \,\partial_{\zeta^-}\bar{A}, \xx
\delta_{\alpha} \bar{F}&=& 
i\,\alpha_1\,\partial_{\xi^-}\psi_R \,-\, i 
\alpha_2\,\partial_{\zeta^-}\psi_L.
\eea
Notice that the supersymmetry transformations given in eqns. 
\eq{susy} and \eq{susybar} are the same as in the 
usual $C=0$ theory.
Also, we have already pointed out the equivalence 
of the $C=0$ part
of our action in eqn. \eq{Ic0p} and the standard  
action~\cite{Braaten:is} given
in eqn. \eq{Iczeroprime}. Hence, the supersymmetry of the  $C=0$ 
part of the action \eq{Ic0p}, is obvious. 
However, one can also explicitly 
check the supersymmetry by varying the action
\eq{Ic0p}, with respect to the supersymmetry 
parameters $\alpha_1$ and $\alpha_2$ and 
using the 
supersymmetry transformations 
\eq{susy} and \eq{susybar}. Thus, the variation of
the action in eqn. \eq{Ic0p} gives:
\be \label{Ic0susy}
\delta I_0 = \int d^2x 
\Bigl[\,I^1_0 \,+\,I^2_0 \,+\, I^3_0 \,+\,I^4_0 \,\Bigr],
\ee
where $\,I^1_0 \,,\,I^2_0 \,,\, I^3_0 \,,\,I^4_0 $
are defined below as:
\bea \label{I01}
I^1_0&& = \Bigl[\,
 \partial_{\xi^-} \partial_{\zeta^-}\bar{A}
\,(\delta A) 
- i \partial_{\zeta^-}(\delta \psi_L) \bar{\psi}_L
- i \partial_{\zeta^-}\psi_L (\delta \bar{\psi}_L) \xx
&&~~~~- i \partial_{\xi^-}(\delta \psi_R) \bar{\psi}_R
- i \partial_{\xi^-}\psi_R (\delta \bar{\psi}_R)
- F (\delta \bar{F})
\,\Bigr]\,
\frac{\partial^2{\mathcal K}} {\partial S\partial\bar{S}},\\
\label{I02}
I^2_0&& = -\Bigl[\,
\Bigl(
  i \partial_{\zeta^-}\psi_L \bar{\psi}_L
+ i \partial_{\xi^-}\psi_R \bar{\psi}_R
+ F\bar{F}
\,\Bigr) \,(\delta A)  \xx
&&~~~~ + i(\delta \bar{\psi}_R)\bar{\psi}_L \bar{F} 
+ i\bar{\psi}_R(\delta \bar{\psi}_L) \bar{F}
+ i\bar{\psi}_R\bar{\psi}_L (\delta \bar{F})
\,\Bigr]\,
\frac{\partial^3{\mathcal K}} {\partial S^2\partial\bar{S}},\\
\label{I03}
I^3_0&& = \Bigl[\,
+ i (\delta \psi_L)\psi_R F 
+ i \psi_L(\delta \psi_R) F
+ i (\delta \bar{\psi}_L)\psi_L\partial_{\zeta^-}\bar{A} 
+\Bigl(\,
\partial_{\xi^-}\bar{A}\partial_{\zeta^-}\bar{A}
\,\Bigr)(\delta A) \xx
&&~~~~
+ i \bar{\psi}_L(\delta \psi_L)\partial_{\zeta^-}\bar{A}
+ i (\delta\bar{\psi}_R)\psi_R\partial_{\xi^-}\bar{A}
+ i \bar{\psi}_R(\delta\psi_R)\partial_{\xi^-}\bar{A}
\,\Bigr]\,
~\frac{\partial^3{\mathcal K}}{\partial S\partial \bar{S}^2}, \\
\label{I04}
I^4_0&& = -\Bigl[\,
\Bigl(\,
   (\delta\bar{\psi}_L)\psi_L\bar{\psi}_R\psi_R
 + \bar{\psi}_L(\delta\psi_L)\bar{\psi}_R\psi_R 
+ \bar{\psi}_L\psi_L(\delta\bar{\psi}_R)\psi_R 
+ \bar{\psi}_L\psi_L\bar{\psi}_R(\delta\psi_R) 
\,\Bigr)  \xx
&&~~~~-i\Bigl(
~ \psi_L\psi_R F +\bar{\psi}_L\psi_L\partial_{\zeta^-}\bar{A}
+ \bar{\psi}_R\psi_R\partial_{\xi^-}\bar{A} 
\,\Bigr) (\delta A)
\,\Bigr]\,
\frac{\partial^4{\mathcal K}} {\partial S^2\partial\bar{S}^2}.
\eea
Now, one can directly substitute the supersymmetry 
transformations given in eqns. \eq{susy} and \eq{susybar}, in the
variation of the action given in eqn. \eq{Ic0susy}, or 
equivalently 
in eqns. \eq{I01}-\eq{I04}. The result is that,
each of the terms in eqns. \eq{I01}-\eq{I04} vanishes 
individually. Therefore, the
$C=0$ part of the action given in eqn. \eq{Ic0p}
preserves $N=\frac{2}{2}$ supersymmetry. Also,
by writing down transformations
under both $Q$ and  $\bar{Q}$'s, we find that the  
supersymmetry variation corresponding to $Q$'s in the action
$I_0$ cancels out, whereas the $C$ dependent terms generated by
$\bar{Q}'$s add up. Therefore,
the action does not remain invariant under $\bar{Q}$'s.

Now, regarding the non(anti)commutative part of the 
action in eqn. \eq{Icnonzerofull},
we will first check
the supersymmetry of the action \eq{Icnonzero1} 
to first order  in $(\det\,C)$. 
The calculations 
get simplified by noting that the
supersymmetry transformations of the 
component fields $F$ and $\bar{A}$ 
vanish identically. Thus, 
varying the action \eq{Icnonzero1}, with respect to the 
parameters $\alpha_1$ and $\alpha_2$, and leaving out the 
terms which cancel out trivially, we end up with:
\be \label{Icsusy}
\delta I_{C1} =  \left(\frac{1}{4}\det\,C \right)\:
\int d^2x\:
\Bigl[\,I^1_C \,+\,I^2_C \,+\, I^3_C \,+\,I^4_C \,\Bigr],
\ee
where $I^1_C,\,I^2_C,\,I^3_C,\,I^4_C\,$ are defined below as:
\bea \label{Ic1}
I^1_C&& \!\!\!\!= \Bigl[\,
 \frac{1}{2} F^2 \partial_{\xi^-} \partial_{\zeta^-}\bar{A}
 \,(\delta A) +\frac{i}{6}\,
\Bigl( -2 F (\delta \bar{\psi}_L)\bar{\psi}_R \partial_{\xi^-} 
 \partial_{\zeta^-}\bar{A} -2 F\bar{\psi}_L(\delta \bar{\psi}_R)
 \partial_{\xi^-} \partial_{\zeta^-}\bar{A} \xx
&& + F^2(\delta\bar{\psi}_L)\partial_{\zeta^-}\psi_L 
 + F^2\bar{\psi}_L\partial_{\zeta^-}(\delta\psi_L)
 +  F^2(\delta\bar{\psi}_R)\partial_{\xi^-}\psi_R \xx
&& +  F^2\bar{\psi}_R\partial_{\xi^-}(\delta\psi_R)
   +i  F^3\delta\bar{F} 
\,\Bigr)
\,\Bigr]\,
\frac{\partial^4{\mathcal K}} {\partial S^3\partial\bar{S}},\\
\label{Ic2}
I^2_C&&\!\!\!\!= \left[\:
\frac{1}{6}\,F\Bigl\{\,
-2i\bar{\psi}_L\bar{\psi}_R\partial_{\xi^-} 
 \partial_{\zeta^-}\bar{A} + 
i\bar{\psi}_L\partial_{\zeta^-}\psi_L F 
 + i\bar{\psi}_R\partial_{\xi^-}\psi_R F - F^2{\bar F}
\,\Bigr\}\,(\delta A)  \right.  \xx
&& \left.+ \frac{i}{6}\, F^2\,
\Bigl\{\, (\delta\bar{\psi}_L)\bar{\psi}_R {\bar F} 
 + \bar{\psi}_L(\delta\bar{\psi}_R) {\bar F}
 + \bar{\psi}_L\bar{\psi}_R (\delta{\bar F})
\,\Bigr\}\,\right]\:
\frac{\partial^5{\mathcal K}} {\partial S^4\partial\bar{S}}, \\
\label{Ic3}
I^3_C &&\!\!\!\!= \left[\,\frac{1}{2}
F^2 \partial_{\xi^-}\bar{A}\partial_{\zeta^-}\bar{A} (\delta A)
 -\frac{i}{6}F
\Bigl\{
   2(\delta\bar{\psi}_L)\bar{\psi}_R \partial_{\xi^-}
   \bar{A}\partial_{\zeta^-}\bar{A}
 + 2\bar{\psi}_L(\delta\bar{\psi}_R) \partial_{\xi^-}
   \bar{A}\partial_{\zeta^-}\bar{A} \right.\xx
&& \left.-F^2(\delta\psi_L)\psi_R
 - F^2\psi_L(\delta\psi_R)  
 + F (\delta\psi_L)\bar{\psi}_L \partial_{\zeta^-}\bar{A}  
 + F \psi_L(\delta\bar{\psi}_L) \partial_{\zeta^-}\bar{A} \right. \xx
&& \left.+ F (\delta\psi_R)\bar{\psi}_R \partial_{\xi^-}\bar{A}
 + F \psi_R(\delta\bar{\psi}_R) \partial_{\xi^-}\bar{A}
\,\Bigr\}
\:\right]\:
\frac{\partial^5{\mathcal K}} {\partial S^3\partial\bar{S}^2} \\
\label{Ic4}
I^4_C &&\!\!\!\!= -\Bigl[\,
\frac{i}{6}F~
\Bigl(~2\,
\bar{\psi}_L\bar{\psi}_R \partial_{\xi^-}
\bar{A}\partial_{\zeta^-}\bar{A} - F^2\psi_L\psi_R 
- F \psi_L\bar{\psi}_L
 \partial_{\zeta^-}\bar{A} 
- F \psi_R\bar{\psi}_R \partial_{\xi^-}\bar{A}
~\Bigr)(\delta A) \xx
&&+\frac{1}{6}
\Bigl(\,
   F^2(\delta\bar{\psi}_L)\psi_L\bar{\psi}_R\psi_R
 + F^2\bar{\psi}_L(\delta\psi_L)\bar{\psi}_R\psi_R \xx
&& + F^2\bar{\psi}_L\psi_L(\delta\bar{\psi}_R)\psi_R
 + F^2\bar{\psi}_L\psi_L\bar{\psi}_R(\delta\psi_R)
\,\Bigr) 
\,\Bigr]\,
\frac{\partial^6{\mathcal K}} {\partial S^4\partial\bar{S}^2}
\eea
Now, one can directly
substitute the supersymmetry transformations of the 
component fields given in \eq{susy} and \eq{susybar}, in the
variation of the action given in eqn. \eq{Icsusy} 
or equivalently in
eqns. \eq{Ic1}-\eq{Ic4}. The outcome being that, each of the
terms in the action given in eqns. \eq{Ic1}-\eq{Ic4}, vanish
individually. Thus, we have the result that the 
non(anti)commutative  part of the 
action given in eqn. \eq{Icnonzero1},
preserves $N=\frac{2}{2}$ supersymmetry to first order in 
$(\det\,C)$. 

To check whether the 
non(anti)commutative  part of the full
action given in eqn. (\ref{Icnonzerofull}) 
preserves $N=\frac{2}{2}$ supersymmetry to all 
orders in $(\det\,C)$,
we perform an explicit verification of supersymmetry. We follow
the procedure as discussed above and vary the action
$I_C$ given in eqn. (\ref{Icnonzerofull}) with respect to the
supersymmetry parameters $\alpha_1$ and $\alpha_2$. 
It turns out that the supersymmetry 
variation of $I^n_{C}$ again consists of four 
parts (ignoring terms cancelling trivially):
\be \label{Icnsusy}
\delta I^n_C =  \left(\frac{1}{4}\det\,C \right)^n\:
\int d^2x\: F^{2n-2}
\Bigl[\,I^1_{Cn} \,+\,I^2_{Cn} \,+\, I^3_{Cn} \,+\,I^4_{Cn}
\,\Bigr],
\ee
where $I^1_{Cn},\,I^2_{Cn},\,I^3_{Cn},\,I^4_{Cn}\,$ 
are defined below as:
\bea \label{Icn1}
I^1_{Cn}&& \!\!\!\!= \Bigl[\,
 \frac{1}{(2n)!} F^2 \partial_{\xi^-} \partial_{\zeta^-}\bar{A}
 \,(\delta A) +\frac{i}{(2n+1)!}\,
\Bigl( -2n F (\delta \bar{\psi}_L)\bar{\psi}_R \partial_{\xi^-} 
 \partial_{\zeta^-}\bar{A} \xx
&&-2n F\bar{\psi}_L(\delta \bar{\psi}_R)
 \partial_{\xi^-} \partial_{\zeta^-}\bar{A} \xx
&& + F^2(\delta\bar{\psi}_L)\partial_{\zeta^-}\psi_L 
 + F^2\bar{\psi}_L\partial_{\zeta^-}(\delta\psi_L)
 +  F^2(\delta\bar{\psi}_R)\partial_{\xi^-}\psi_R \xx
&& +  F^2\bar{\psi}_R\partial_{\xi^-}(\delta\psi_R)
   +i  F^3\delta\bar{F} 
\,\Bigr)
\,\Bigr]\,
\frac{\partial^{2n+2}{\mathcal K}} 
{\partial S^{2n+1}\partial\bar{S}},\\
\label{Icn2}
I^2_{Cn}&&\!\!\!\!= \frac{F}{(2n+1)!}\Bigl[\,
\Bigl
 (-2i\,n\,\bar{\psi}_L\bar{\psi}_R\partial_{\xi^-} 
 \partial_{\zeta^-}\bar{A} 
+ i\bar{\psi}_L\partial_{\zeta^-}\psi_L F 
 + i\bar{\psi}_R\partial_{\xi^-}\psi_R F - F^2{\bar F}
\Bigr)(\delta A) \xx
&& + i(\delta\bar{\psi}_L)\bar{\psi}_R F{\bar F} 
 + i\bar{\psi}_L(\delta\bar{\psi}_R) F{\bar F}
 + i\bar{\psi}_L\bar{\psi}_R F(\delta{\bar F})
\,\Bigr]\,
\frac{\partial^{2n+3}{\mathcal K}} 
{\partial S^{2n+2}\partial\bar{S}}, \\
\label{Icn3}
I^3_{Cn} &&\!\!\!\!= \Bigl[\,\frac{1}{(2n)!}
F^2 \partial_{\xi^-}\bar{A}\partial_{\zeta^-}\bar{A} (\delta A)
 -\frac{i}{(2n+1)!}\,F \,
\Bigl\{
   2\,n\,(\delta\bar{\psi}_L)\bar{\psi}_R \partial_{\xi^-}
   \bar{A}\partial_{\zeta^-}\bar{A} \xx
&& + 2\,n\,\bar{\psi}_L(\delta\bar{\psi}_R) \partial_{\xi^-}
   \bar{A}\partial_{\zeta^-}\bar{A} \xx
&& - F^2(\delta\psi_L)\psi_R
 - F^2\psi_L(\delta\psi_R)  
 + F (\delta\psi_L)\bar{\psi}_L \partial_{\zeta^-}\bar{A}  
 + F \psi_L(\delta\bar{\psi}_L) \partial_{\zeta^-}\bar{A} \xx
&& + F (\delta\psi_R)\bar{\psi}_R \partial_{\xi^-}\bar{A}
 +F \psi_R(\delta\bar{\psi}_R) \partial_{\xi^-}\bar{A}
\,\Bigr\}\,
\,\Bigr]\,
\frac{\partial^{2n+3}{\mathcal K}} 
{\partial S^{2n+1}\partial\bar{S}^{2}} \\
\label{Icn4}
I^4_{Cn} &&\!\!\!\!= -\,\frac{1}{(2n+1)!}\,\Bigl[\,i
F~
\Bigl(~2n
\bar{\psi}_L\bar{\psi}_R \partial_{\xi^-}
\bar{A}\partial_{\zeta^-}\bar{A} - F^2\psi_L\psi_R 
- F \psi_L\bar{\psi}_L
 \partial_{\zeta^-}\bar{A} \xx
&&- F \psi_R\bar{\psi}_R \partial_{\xi^-}\bar{A}
~\Bigr)(\delta A) 
+
\Bigl(\,
   F^2(\delta\bar{\psi}_L)\psi_L\bar{\psi}_R\psi_R
 + F^2\bar{\psi}_L(\delta\psi_L)\bar{\psi}_R\psi_R \xx
&& + F^2\bar{\psi}_L\psi_L(\delta\bar{\psi}_R)\psi_R
 + F^2\bar{\psi}_L\psi_L\bar{\psi}_R(\delta\psi_R)
\,\Bigr) 
\,\Bigr]\,
\frac{\partial^{2n+4}{\mathcal K}} 
{\partial S^{2n+2}\partial\bar{S}^{2}}
\eea
Now, one can go on to substitute the supersymmetry variation of
the component fields given in eqns. (\ref{susy}) and 
(\ref{susybar}) and hence in eqns. (\ref{Icn1})-(\ref{Icn4}).
Owing to some beautiful cancellations, all the pieces of the 
supersymmetry variation of the full action
$I_C$ given in eqn. (\ref{Icnsusy}) vanish identically. Hence,
the non(anti)commutative part of the supersymmetric theory
given in eqn. (\ref{Icnonzerofull}) preserves
$N=\frac{2}{2}$ supersymmetry to all orders in 
$(\det\,C)$. 

Combining the results obtained for the supersymmetry variations
of the actions given in eqns. (\ref{Ic0p}) and 
(\ref{Icnonzerofull}), we conclude that the full 
non(anti)commutative theory given by eqn. \eq{break},  
preserves  $N=\frac{2}{2}$ supersymmetry.

\subsection{\bf{Superpotential }}           

In this section, we generalize the results of the preceding
section, by including superpotentials in the action \eq{action}.
The complete kinetic action for the non(anti)commutative
$N=\frac{2}{2}$ theory was given in eqn. \eq{break}.
The action can be generalized to include arbitrary 
superpotentials, as shown :
\be
I_g = \int d^2x \:d\theta \:d\chi\: d\bar{\theta}\: d\bar{\chi}\:
{\mathcal K}(S,\bar{S})
+ \int d^2x \:d\theta \:d\chi\: W(S)
+ \int d^2x\: d\bar{\theta}\: d\bar{\chi}\: \bar{W}(\bar{S}),
\ee
where apart from the K\"{a}hler potential ${\mathcal K}(S,\bar{S})$,
we also have superpotentials $W(S)$ and 
$\bar{W}(\bar{S})$. Like the case for  ${\mathcal K}(S,\bar{S})$
in eqn. \eq{expand}, we expand the superpotentials
in terms of the component fields as shown:
\bea \label{W}
&&W(S) = W(A) \,+\, L\: \frac{\partial W}{\partial S} |_{S=A} \,+ 
\, \frac{1}{2!}\,L*L\,\frac{\partial^2 W}{\partial S^2} |_{S=A}\,
+\, \frac{1}{3!} \,L*L*L\,
\frac{\partial^3W}{\partial S^3} |_{S=A} \xx
&&~~~~~\,+\,\cdots\,+\,  \frac{1}{(2n)!}\: L_*^{2n}\: 
\frac{\partial^{2n}W}{\partial S^{2n}} |_{S=A}\,+\,\cdots
\,+\,\frac{1}{(2n+1)!}\:L_*^{2n+1}\: 
\frac{\partial^{2n+1}W}{\partial S^{2n+1}}|_{S=A}\,+\,\cdots\,.
\eea
Note that the quantities $L,\, L_*^{2n}$ and $L_*^{2n+1}$ 
have been defined in eqns. \eq{LRnota}, \eq{Lstar} and \eq{L1star} 
respectively. We also have:
\bea \label{Wbar}
&&\bar{W}(\bar{S}) = 
\bar{W}(\bar{A}) \,+\, R\: 
\frac{\partial \bar{W}}{\partial \bar{S}} |_{\bar{S}=\bar{A}} \,+ 
\, \frac{1}{2!}\,R*R\,
\frac{\partial^2 \bar{W}}{\partial S^2} |_{\bar{S}=\bar{A}},
\eea
where $R$ appearing above, has been defined in eqn. \eq{LRnota}. 
Notice that the above expansion of $\bar{W}(\bar{S})$ 
gets truncated, since $R_*^n = 0,\: {\rm for}\:n\,\geq 3$.
Hence, using the results of section-3, 
we can evaluate various terms in the superpotential in 
eqn. (\ref{W}). Collecting terms proportional to
$\theta \,\chi$ which will ultimately contribute to the action, 
we end up with:
\bea \label{Wexp}
&&W(S)|_{\theta\,\chi\,}
 = -\:i F\:\frac{\partial W}{\partial S} |_{S=A} \,-
\,\bar{\psi}_L\bar{\psi}_R \,
\frac{\partial^2 W}{\partial S^2} |_{S=A}\, \xx
&&~~~~~~~~~~- \sum_{n=2}^{\infty}
\,\frac{1}{(2n-1)!}\: 
\left(\,\frac{1}{4} \det~C\right)^{n-1} \,F^{2n-2}
\,\:\bar{\psi}_L\,\bar{\psi}_R \: 
\frac{\partial^{2n}W}{\partial S^{2n}} |_{S=A}\,\xx
&&~~~~~~~~~~- \sum_{n=1}^{\infty}
\,\frac{i}{(2n+1)!}\: 
\left(\,\frac{1}{4} \det~C\right)^{n} \,F^{2n+1}
\frac{\partial^{2n+1}W}{\partial 
S^{2n+1}} |_{S=A}\,+\,\cdots\,.
\eea
Similarly, one can also write down the
terms coming from eqn. (\ref{Wbar}). 
We end up with
the following most general form 
of $\,\bar{W}(\bar{S})\,$, after collecting
the terms proportional to $\bar{\theta} \,\bar{\chi}$ :
\bea
&&\bar{W}(\bar{S})|_{\bar{\theta}
\,\bar{\chi}\,}
=  -\:i\bar{F}\:
\frac{\partial \bar{W}}{\partial {\bar S}}|_{\bar{S}=\bar{A}}\,
-\,\psi_L\psi_R \: 
\frac{\partial^2 \bar{W}}{\partial \bar{S}^2} |_{\bar{S}=\bar{A}}.
\eea

\section{\bf{Discussion}}           

To conclude, we have analyzed the most general
classical action of an 
$ {\cal N}=2$ supersymmetric theory with single chiral and 
antichiral superfield,
defined on a 
non(anti)commutative superspace, for an arbitrary K\"{a}hler
and super potential. The key aspect of the analysis is 
the emergence of a
power series expansion in  $(\det\,C)$ in the action.
Writing down a general  $n^{\rm th}$ order term in the
action is possible and is found to be Lorentz
invariant. 
The $N=\frac{2}{2}$ supersymmetry of the action
was explicitly shown, order by order 
in $(\det\,C)$. 

The $C=0$ part of the 
supersymmetric theory studied in this paper, has a 
straightforward generalization to supersymmetric
nonlinear $\sigma$-model containing several superfields. 
From the definitions of 
geometric quantities given in section-2.2, the 
action $I_0$ given in eqn. \eq{Iczero},
can be written as~\cite{Braaten:is}:
\bea  \label{icgeo}
I_0 &=& \Bigl[\,\int d^2x ~g_{i\bar{j}}~\Bigl(~-\partial_{\xi^-}A^{i}
\partial_{\zeta^-}\bar{A}^{j}
+i\bar{\psi}^{i}_L\partial_{\zeta^-}\psi^{j}_L
+ i\bar{\psi}^{i}_R\partial_{\xi^-}\psi^{j}_R 
- F^{i}{\bar F}^{j}  ~\Bigr) \xx
&&~~~~~~~+ \Gamma_{ij{\bar k}}
\left(~\bar{\psi}^{i}_L\bar{\psi}^{j}_R{\bar F}^{k}~\right)~  
+ i\Gamma_{\bar{i}\bar{j}k}\left(~\psi^{i}_L\psi^{j}_R F^{k} 
-\bar{\psi}^{k}_L\psi^{i}_L\partial_{\zeta^-}\bar{A}^{j}
-  \bar{\psi}^{k}_R\psi^{i}_R\partial_{\xi^-}\bar{A}^{j}~\right) \xx
&&~~~~~~~~- R_{{\bar i}j{\bar k}l}
\left(~\psi^{i}_L \bar{\psi}^{j}_L\psi^{k}_R\bar{\psi}^{l}_R 
 ~\right)\,\Bigr].
\eea
It seems plausible 
that the action for the complete $N=\frac{2}{2}$
supersymmetric theory with arbitrary number of superfields 
$S^i$ and $\bar{S}^i$ can also 
be written in terms of the  K\"{a}hler metric and the
subsequent geometric quantities following from it. 
However, to 
geometrize the full $N=\frac{2}{2}$ supersymmetric theory
studied in this paper, the
identities given eqns. \eq{christo} and \eq{curv}, 
are not enough. Also terms of the type 
$\psi^i_L\cdots\psi^n_R$
etc., will appear in the expressions such as $L_*^n$. It 
will be interesting to examine whether the complete 
action can be written in a geometric form like (\ref{icgeo}).

There are lots of other avenues one can explore. 
One of the most obvious
things to study is the quantum aspects of the 
non(anti)commutative theory we have discussed. It is a well
acclaimed fact that the K\"{a}hler geometry imposes 
severe restrictions on the kind of counter terms which
can appear in the quantum action. In view of the 
series expansion we see in the classical action, it is
worthwhile to analyze what kind of restrictions  
put by  K\"{a}hler geometry appear at the 
quantum level. Moreover, checking the 
renormalizability of the
above model is an interesting aspect to be pursued.

Further, in two dimensions one can have new
$\sigma$-models due to the possibility of defining
twisted multiplets~\cite{Gates:nk}, 
in addition to the chiral and antichiral multiplets.
These $\sigma$-models have established various 
connections
between $D=4,\: {\cal N} =1$ and $D=2,\:  {\cal N}=2$ models. 
It may be possible to generalize the result of this 
paper, to include twisted (chiral and antichiral) 
fields as well. In this way, 
one can possibly study the consequences of Mirror symmetry 
in $D=2,\: N=(2,2)$ models, which interchanges 
the (anti)chiral and twisted (anti)chiral fields.

\appendix
\section*{{Appendix A.}}
\renewcommand{\theequation}{A-\arabic{equation}}
\setcounter{equation}{0}

Here, we provide the proof of the identity:
\be \label{lrlrl}
L_*^{m}\,*R\,*\,L_*^{n}\,*\,R\,*\,L_*^{p}\,
|_{\bar{\theta}
\,\bar{\chi}\,\theta\,\chi\,}
= L_*^{(m+n+p)}\,*R_*^2\,|_{\bar{\theta}
\,\bar{\chi}\,\theta\,\chi\,}\,,
\ee
where $m,n$ and $p$ are integers.
As already mentioned, the above identity strictly holds 
only for terms with coefficient 
$\,\bar{\theta}\,\bar{\chi}\,(\theta\,\chi)\,$. Before
proceeding, below we collect certain results from the text,
which will be used repeatedly:
\bea \label{La}
L&& =~ + i\theta\,\bar{\psi}_L - i\chi\,\bar{\psi}_R 
   - i(\theta\,\chi)\, F, \\
\label{Ra}
R&& =~- i\bar{\theta}\,\psi_L 
 + i\bar{\chi}\psi_R + i \theta\bar{\theta}
\partial_{\xi^-}\bar{A} + i\chi\bar{\chi} 
\partial_{\zeta^-}\bar{A} \xx
&&~~~+~\bar{\theta}\,\bar{\chi}\left(-\chi\,\partial_{\zeta^-}\psi_L
-\theta\,\partial_{\xi^-}\psi_R -i\bar{F}+
(\theta\,\chi)\,\partial_{\xi^-}
\partial_{\zeta^-} \bar{A}  \right),\\
\label{Lstara}
L_*^{2n}&&=~ \left(\,\frac{1}{4} \det~C\right)^{n-1} \,F^{2n-2}
\,\left[-2n \,(\theta \chi)\:
\bar{\psi}_L\,\bar{\psi}_R \,+\,  
\left(\frac{1}{4}\det~C\right)\, F^2\, \right], \\
\label{L1stara}
L_*^{2n+1}&&=~\left(\,\frac{1}{4} \det~C\right)^{n}\, F^{2n-1}\,
\left(-2n\, i\,\bar{\psi}_L\,\bar{\psi}_R \,+\,  F.L\,\right),\\
\label{R2stara}
R_*^2 &&=~ 2 \bar{\theta}\,\bar{\chi}\,
\Bigl( -\psi_L\psi_R \,-\,\chi\,\psi_L
  \,\partial_{\zeta^-} \bar{A}\,
  -\,\theta\,\psi_R\,\partial_{\xi^-} \bar{A}\,         
+\,(\theta\,\chi)\,\partial_{\xi^-} \bar{A}\partial_{\zeta^-} \bar{A}
\Bigr),  \\
\label{Rnstara}
R_*^{n}&&=~ 0,\quad {\rm for}~~ n \,>\, 2.
\eea

To begin, using the  
definitions of $L$ and $R$ given in eqns. (\ref{La}),
(\ref{Ra}) we deduce the following result:
\bea \label{lr}
R\,*\,L  &&= \frac{1}{2}\bar{\theta}\,
\Bigl(C^{00}\,\bar{\psi}_L\,\partial_{\xi^-}\bar{A}
+C^{01}\,\bar{\psi}_R\,\partial_{\xi^-}\bar{A}\Bigr)
-\frac{1}{2}\bar{\chi}\,
\Bigl(C^{01}\,\bar{\psi}_L\,\partial_{\zeta^-}\bar{A} 
+C^{11}\,\bar{\psi}_R\,\partial_{\zeta^-}\bar{A} \Bigr) \xx
&&~~-\,\bar{\chi}\,\theta\,
\Bigl(\,\bar{\psi}_L\,\psi_R \, 
- \frac{1}{2}\,C^{11}\,F\,\partial_{\zeta^-}\bar{A}
\Bigr)
-\,\bar{\theta}\,\chi\,
\Bigl(\,\bar{\psi}_R\,\psi_L \,
+ \frac{1}{2}\,C^{00}\,F\,\partial_{\xi^-}\bar{A}
\Bigr) \xx
&&~~-\,\bar{\theta}\,\theta\,
\Bigl(\,-\bar{\psi}_L\,\psi_L \,
+ \frac{1}{2}\,C^{01}\,F\,\partial_{\xi^-}\bar{A}
\Bigr)
+\,\bar{\chi}\,\chi\,
\Bigl(\,\bar{\psi}_R\,\psi_R \,  
+ \frac{1}{2}\,C^{10}\,F\,\partial_{\zeta^-}\bar{A}
\Bigr) \xx
&&~~+\,\frac{i}{2}\: \bar{\theta}\,\bar{\chi}\,
\Bigl(\, C^{10} \,\bar{\psi}_L  \partial_{\zeta^-}\psi_L
+ \, C^{11} \,\bar{\psi}_R  \partial_{\zeta^-}\psi_L 
-\, C^{00} \,\bar{\psi}_L  \partial_{\xi^-}\psi_R
-\, C^{01} \,\bar{\psi}_R  \partial_{\xi^-}\psi_R \xx
&&~~+ \frac{1}{2}
  (\det~C)\,F\,\partial_{\xi^-}\partial_{\zeta^-}\bar{A} 
\,\Bigr)         
-\,\bar{\theta}\,(\theta\,\chi)\,
\left(\bar{\psi}_R\,
\partial_{\xi^-}\bar{A} \,+\,F\,\psi_L \,\right)\, \xx
&&~~-\,\bar{\chi}\,(\theta\,\chi)\,
\left(\bar{\psi}_L\,
\partial_{\zeta^-}\bar{A} \,-\,F\,\psi_R\right)\,
+\,\bar{\theta}\,\bar{\chi}\,\theta\,
\Bigl\{ \frac{i}{2}\,
\Bigl(\,- C^{11} \,\partial_{\zeta^-}\psi_L \,F 
+ \, C^{10} \,\partial_{\xi^-}\psi_R\,F \xx
&&~~-\, C^{01} \,\bar{\psi}_L  
          \partial_{\xi^-}\partial_{\zeta^-}\bar{A}
-\, C^{11} \,\bar{\psi}_R  
          \partial_{\xi^-}\partial_{\zeta^-}\bar{A}
\,\Bigr\}\,
+\,\bar{\psi}_L\,\bar{F}\,
\,\Bigr)        
+\,\bar{\theta}\,\bar{\chi}\,\chi\,
\Bigl\{ \frac{i}{2}\,
\Bigl(\, C^{00} \,\partial_{\xi^-}\psi_R \,F \xx
&&~~- \, C^{10} \,\partial_{\zeta^-}\psi_L\,F
-\, C^{01} \,\bar{\psi}_R  
          \partial_{\xi^-}\partial_{\zeta^-}\bar{A} 
-\, C^{00} \,\bar{\psi}_L  
          \partial_{\xi^-}\partial_{\zeta^-}\bar{A}
\,\Bigr\}\,
-\,\bar{\psi}_R\,\bar{F}\,
\,\Bigr) \xx
&&~~+\,\bar{\theta}\,\bar{\chi}\,(\theta\,\chi)\,
\left( i\,\bar{\psi}_L\,\partial_{\zeta^-}\psi_L
\,+\, i\,\bar{\psi}_R\,
\partial_{\xi^-}\psi_R \,-\, F\,{\bar F}\right).
\eea
One can similarly obtain $L*R$ as given below: 
\bea \label{rl}
L\,*\,R  &&= -\,\frac{1}{2}\,\bar{\theta}\,
\Bigl(C^{00}\,\bar{\psi}_L\,\partial_{\xi^-}\bar{A}
+C^{01}\,\bar{\psi}_R\,\partial_{\xi^-}\bar{A}\Bigr)
+\,\frac{1}{2}\,\bar{\chi}\,
\Bigl(C^{01}\,\bar{\psi}_L\,\partial_{\zeta^-}\bar{A} 
+C^{11}\,\bar{\psi}_R\,\partial_{\zeta^-}\bar{A} \Bigr) \xx
&&~~-\,\bar{\chi}\,\theta\,
\Bigl(\,\bar{\psi}_L\,\psi_R \, 
+ \frac{1}{2}\,C^{11}\,F\,\partial_{\zeta^-}\bar{A}
\Bigr)
-\,\bar{\theta}\,\chi\,
\Bigl(\,\bar{\psi}_R\,\psi_L \,
- \frac{1}{2}\,C^{00}\,F\,\partial_{\xi^-}\bar{A}
\Bigr) \xx
&&~~-\,\bar{\theta}\,\theta\,
\Bigl(\,-\bar{\psi}_L\,\psi_L \,
- \frac{1}{2}\,C^{01}\,F\,\partial_{\xi^-}\bar{A}
\Bigr)
+\,\bar{\chi}\,\chi\,
\Bigl(\,\bar{\psi}_R\,\psi_R \,  
- \frac{1}{2}\,C^{10}\,F\,\partial_{\zeta^-}\bar{A}
\Bigr) \xx
&&~~-\,\frac{i}{2}\: \bar{\theta}\,\bar{\chi}\,
\Bigl(\, C^{10} \,\bar{\psi}_L  \partial_{\zeta^-}\psi_L
+ \, C^{11} \,\bar{\psi}_R  \partial_{\zeta^-}\psi_L 
-\, C^{00} \,\bar{\psi}_L  \partial_{\xi^-}\psi_R
-\, C^{01} \,\bar{\psi}_R  \partial_{\xi^-}\psi_R \xx
&&~~- \frac{1}{2}
  (\det~C)\,F\,\partial_{\xi^-}\partial_{\zeta^-}\bar{A} 
\,\Bigr)         
-\,\bar{\theta}\,(\theta\,\chi)\,
\left(\bar{\psi}_R\,
\partial_{\xi^-}\bar{A} \,+\,F\,\psi_L \,\right)\, \xx
&&~~-\,\bar{\chi}\,(\theta\,\chi)\,
\left(\bar{\psi}_L\,
\partial_{\zeta^-}\bar{A} \,-\,F\,\psi_R\right)\,
-\,\bar{\theta}\,\bar{\chi}\,\theta\,
\Bigl\{ \frac{i}{2}\,
\Bigl(\,- C^{11} \,\partial_{\zeta^-}\psi_L \,F 
+ \, C^{10} \,\partial_{\xi^-}\psi_R\,F \xx
&&~~-\, C^{01} \,\bar{\psi}_L  
          \partial_{\xi^-}\partial_{\zeta^-}\bar{A}
-\, C^{11} \,\bar{\psi}_R  
          \partial_{\xi^-}\partial_{\zeta^-}\bar{A}
\,\Bigr\}\,
-\,\bar{\psi}_L\,\bar{F}\,
\,\Bigr)        
-\,\bar{\theta}\,\bar{\chi}\,\chi\,
\Bigl\{ \frac{i}{2}\,
\Bigl(\, C^{00} \,\partial_{\xi^-}\psi_R \,F \xx
&&~~- \, C^{10} \,\partial_{\zeta^-}\psi_L\,F
-\, C^{01} \,\bar{\psi}_R  
          \partial_{\xi^-}\partial_{\zeta^-}\bar{A} 
-\, C^{00} \,\bar{\psi}_L  
          \partial_{\xi^-}\partial_{\zeta^-}\bar{A}
\,\Bigr\}\,
+\,\bar{\psi}_R\,\bar{F}\,
\,\Bigr) \xx
&&~~+\,\bar{\theta}\,\bar{\chi}\,(\theta\,\chi)\,
\left( i\,\bar{\psi}_L\,\partial_{\zeta^-}\psi_L
\,+\, i\,\bar{\psi}_R\,
\partial_{\xi^-}\psi_R \,-\, F\,{\bar F}\right).
\eea
From eqn. (\ref{lr}),
one can notice that 
all the terms in  $R*L$, which 
depend on $C^{\alpha\beta}$ (Lorentz non invariant) 
come with opposite signs
to the ones in $L*R$ given in eqn. (\ref{rl}). 
However, other terms in $R*L$ 
independent of $C^{\alpha\beta}$  come
with the same sign as in $L*R$. In particular, we
notice that the following results hold true when
we restrict ourselves to terms  which are proportional to 
$\,\bar{\theta}\,\bar{\chi}\,(\theta\,\chi)\,$:
\be \label{lrrl}
L\,*\,R |_{\bar{\theta}
\,\bar{\chi}\,\theta\,\chi\,}  
~~=~~  R\,*\,L|_{\bar{\theta}
\,\bar{\chi}\,\theta\,\chi\,}\,.
\ee
Using eqns. (\ref{lr}) and (\ref{rl}), one can also 
show the following:
\bea
L_*^2\,*\,R |_{\bar{\theta}
\,\bar{\chi}\,\theta\,\chi\,}  
&&\!\!=~~  L\,*\,R\,*\,L|_{\bar{\theta}
\,\bar{\chi}\,\theta\,\chi\,}\,\\
\label{llr}
&&\!\!=~~R\,*\,L_*^2|_{\bar{\theta}
\,\bar{\chi}\,\theta\,\chi\,}\,,\\
 R_*^2\,*\,L|_{\bar{\theta}
\,\bar{\chi}\,\theta\,\chi\,}\, 
&&\!\!=~~ R\,*\,L\,*\,R |_{\bar{\theta}
\,\bar{\chi}\,\theta\,\chi\,}  \\
\label{rrl}
&&\!\!=~~ L\,*\, R_*^2 |_{\bar{\theta}
\,\bar{\chi}\,\theta\,\chi\,} .
\eea
Now, the combination
in which $L*R$ appears in the action is 
$\,[\,L*R\,]\,=\,L*R + R*L$, in which the
terms proportional to $C^{\alpha\beta}$ 
cancel out. 

Further, using the definition of $R$ given in 
eqn. (\ref{Ra}) and the results in 
eqns. (\ref{Lstara})-(\ref{L1stara}), 
the following identities can be proved:
\bea \label{LRf1}
L_*^{2n}\,*\,R \,|_{\bar{\theta}
\,\bar{\chi}\,\theta\,\chi\,}\,&&
=~L\,*\,R\,*\,L_*^{2n-1}|_{\bar{\theta}
\,\bar{\chi}\,\theta\,\chi\,}\,=\cdots
=~ R\,*L_*^{2n}|_{\bar{\theta}
\,\bar{\chi}\,\theta\,\chi\,}\, ,\\
L_*^{2n+1}\,*\,R \,|_{\bar{\theta}
\,\bar{\chi}\,\theta\,\chi\,}\,&&
=~L\,*\,R\,*\,L_*^{2n}|_{\bar{\theta}
\,\bar{\chi}\,\theta\,\chi\,}\, =\cdots
=~ R\,*L_*^{2n+1}|_{\bar{\theta}
\,\bar{\chi}\,\theta\,\chi\,}\, , \\
L_*^{2n}\,*\,R_*^2 \,|_{\bar{\theta}
\,\bar{\chi}\,\theta\,\chi\,}\,&&
=~L_*^{2n-1}\,*R\,*\,L\,*R|_{\bar{\theta}
\,\bar{\chi}\,\theta\,\chi\,}\,=\cdots
=~ R_*^2\,*L_*^{2n}|_{\bar{\theta}
\,\bar{\chi}\,\theta\,\chi\,}\,,  \\ 
\label{LRf3}
L_*^{2n+1}\,*\,R_*^2 \,|_{\bar{\theta}
\,\bar{\chi}\,\theta\,\chi\,}\,&&
=~R\,*\,L_*^{2n}\,*R\,*L|_{\bar{\theta}
\,\bar{\chi}\,\theta\,\chi\,}\,=\cdots
=~ R_*^2\,*L_*^{2n+1}|_{\bar{\theta}
\,\bar{\chi}\,\theta\,\chi\,}\, .
\eea
Below, we give conclusive evidence 
for these identities. 

\begin{enumerate}
\item
First, we show that for terms
with coefficient 
$\,\bar{\theta}\,\bar{\chi}\,(\theta\,\chi)\,$, the 
following result holds true:
\be \label{l2nrsum}
L_*^{2n}\,*\,R \,|_{\bar{\theta}
\,\bar{\chi}\,\theta\,\chi\,}\,~=~
 R\,*L_*^{2n}|_{\bar{\theta}
\,\bar{\chi}\,\theta\,\chi\,}\,.
\ee
From the definition of $R$ given in eqn. (\ref{Ra})
and the result for in eqn. (\ref{Lstara}), we
can calculate the following:
\bea \label{rl2n}
R\,*\,L_*^{2n}&&=  
\left(\frac{1}{4}\det~C\right)^{n-1}\,F^{2n-2} (-2n)
\bar{\psi}_L\,\bar{\psi}_R\,
\Bigl(\,-i(\theta\chi){\bar \theta}\,\psi_L + 
i(\theta\chi){\bar \chi}\,\psi_R  \xx 
&&+\frac{i}{2}\Bigl( C^{00}\chi{\bar \theta} 
+ C^{01}\theta{\bar \theta}\Bigr) 
\partial_{\xi^-}\bar{A}
-\frac{i}{2}\Bigl( C^{11}\theta{\bar \chi} 
+ C^{10}\chi{\bar \chi}\Bigr) 
\partial_{\zeta^-}\bar{A}\,  \xx  
&&+\,\bar{\theta}{\bar \chi}\:
\Big\{ \, \frac{1}{2} \Bigl( C^{11}\theta 
+ C^{10}\chi\Bigr) \partial_{\zeta^-}\psi_L
- \frac{1}{2} \Bigl( C^{01}\theta 
+ C^{00}\chi\Bigr) \partial_{\xi^-}\psi_R
-i(\theta\chi) \bar{F}  \xx
&&- \left(\frac{1}{4}\det~C\right)
\partial_{\xi^-}\partial_{\zeta^-}\bar{A}
\,\Bigr\}
\:\,\Bigr)
\:+\:    
\left(\frac{1}{4}\det~C\right)^n\,F^{2n}\:R \,,
\eea
Similarly, one can also obtain:
\bea \label{l2nr}
L_*^{2n}\,*\,R&&=  
\left(\frac{1}{4}\det~C\right)^{n-1}\,F^{2n-2} (-2n)
\bar{\psi}_L\,\bar{\psi}_R\,
\Bigl(\,-i(\theta\chi){\bar \theta}\,\psi_L + 
i(\theta\chi){\bar \chi}\,\psi_R  \xx 
&&-\frac{i}{2}\Bigl( C^{00}\chi{\bar \theta} 
+ C^{01}\theta{\bar \theta}\Bigr) 
\partial_{\xi^-}\bar{A}
+\frac{i}{2}\Bigl( C^{11}\theta{\bar \chi} 
+ C^{10}\chi{\bar \chi}\Bigr) 
\partial_{\zeta^-}\bar{A}\,  \xx  
&&+\,\bar {\theta}{\bar \chi} \:
\Big\{ \,- \frac{1}{2} \Bigl( C^{11}\theta 
+ C^{10}\chi\Bigr) \partial_{\zeta^-}\psi_L
+ \frac{1}{2} \Bigl( C^{01}\theta 
+ C^{00}\chi\Bigr) \partial_{\xi^-}\psi_R
-i(\theta\chi) \bar{F}  \xx
&&- \left(\frac{1}{4}\det~C\right)
\partial_{\xi^-}\partial_{\zeta^-}\bar{A}
\,\Bigr\}
\:\,\Bigr)
\:+\:
\left(\frac{1}{4}\det~C\right)^n\,F^{2n}\:R \,.
\eea
Comparing the results in equations (\ref{rl2n}) and
(\ref{l2nr}), one can infer that the identity 
given eqn. (\ref{l2nrsum}) holds true. Further, the 
identity in eqn. (\ref{l2nrsum}), 
seems to hold, even for other terms which neither
depend on $C$ nor on 
$\,\bar{\theta}\,\bar{\chi}\,(\theta\,\chi)\,$.
However, these will not contribute to the 
action after integration over the Grassmannian variables,
and hence, are not interesting for our purposes.
Now, from the identity (\ref{l2nrsum}) and 
using eqns. (\ref{rl2n}) and (\ref{l2nr})
it is possible to see that the following 
relations also hold true:
\bea \label{rl2nr}
\,L_*^{2n}\,*\,R_*^2 |_{\bar{\theta}
\,\bar{\chi}\,\theta\,\chi\,}\,
&&\!\!=~~ \,
R\,*L_*^{2n}\,*R\,|_{\bar{\theta}
\,\bar{\chi}\,\theta\,\chi\,}\,\xx
&&\!\!=~~ R_*^2\,*\,L_*^{2n}|_{\bar{\theta}
\,\bar{\chi}\,\theta\,\chi\,}\,,\\
\label{ln1r}
L_*^{2n+1}\,*R\,|_{\bar{\theta}
\,\bar{\chi}\,\theta\,\chi\,}\,
&&\!\!=~~ L\,*\,R\,*\,L_*^{2n}|_{\bar{\theta}
\,\bar{\chi}\,\theta\,\chi\,}\,. 
\eea

\item
Next, making
use of the result given in eqn. (\ref{L1stara}), one
can directly compute the following quantities:
\bea
\label{Rln1}
R\,*\,L_*^{2n+1}
&&=  
\left(\frac{1}{4}\det~C\right)^n\,F^{2n-1}
\Bigl[\, -2n\,i\,\bar{\psi}_L\,\bar{\psi}_R\,R 
\,+\,F\,\Bigl(\,R\,*\,L
\,\Bigr)
\:\Bigr] \\
\label{ln1R}
L_*^{2n+1}\,*R
&&=  
\left(\frac{1}{4}\det~C\right)^n\,F^{2n-1}
\Bigl[\, -2n\,i\,\bar{\psi}_L\,\bar{\psi}_R\,R 
\,+\,F\,\Bigl(\, L\,*R
\,\Bigr)
\:\Bigr].
\eea
In general, the quantities in  
eqns. (\ref{Rln1}) and (\ref{ln1R}) are not equivalent.
However, for terms with coefficient 
${\bar{\theta}\,\bar{\chi}\,(\theta\,\chi)\,}$, the 
result in eqn. (\ref{lrrl}) can be used to see the 
equivalence of second terms and hence, the 
following identity can be established:
\be \label{L2n1R}
L_*^{2n+1}\,*R\,|_{\bar{\theta}
\,\bar{\chi}\,\theta\,\chi\,}\, 
~~=~~R\,*\,L_*^{2n+1}\,|_{\bar{\theta}
\,\bar{\chi}\,\theta\,\chi\,}\,.
\ee
\item
Similarly, from eqn. (\ref{L1stara}) one can
also calculate the following quantities:
\bea
\label{R2l2n1}
R_*^2\,*\,L_*^{2n+1}
&&=  
\left(\frac{1}{4}\det~C\right)^n\,F^{2n-1}
\Bigl[\, -2n\,i\,\bar{\psi}_L\,\bar{\psi}_R\,R_*^2 
\,+\,F\,\Bigl(\,R_*^2\,*\,L
\,\Bigr)
\:\Bigr] \\
\label{l2n1R2}
L_*^{2n+1}\,*R_*^2
&&=  
\left(\frac{1}{4}\det~C\right)^n\,F^{2n-1}
\Bigl[\, -2n\,i\,\bar{\psi}_L\,\bar{\psi}_R\,R_*^2 
\,+\,F\,\Bigl(\, L\,*R_*^2
\,\Bigr)
\:\Bigr].
\eea
Now, using the result given in eqn. (\ref{rrl}) in 
eqn. (\ref{R2l2n1}), the 
following equivalence can be shown 
for terms with coefficient 
${\bar{\theta}\,\bar{\chi}\,(\theta\,\chi)\,}$ :
\be
\label{R2ln1}
R_*^2\,*\,L_*^{2n+1}|_{\bar{\theta}
\,\bar{\chi}\,\theta\,\chi\,}\,
~~=~~L_*^{2n+1}\,*\,R_*^2|_{\bar{\theta}
\,\bar{\chi}\,\theta\,\chi\,}\,.
\ee
Moreover, from eqns. (\ref{Rln1}) and  (\ref{L2n1R}),
one can also derive the following identities, when
restricted to terms with coefficient
${\bar{\theta}\,\bar{\chi}\,(\theta\,\chi)\,}$ :
\bea \label{r2l2n1}
R_*^2\,*\,L_*^{2n+1}|_{\bar{\theta}
\,\bar{\chi}\,\theta\,\chi\,}\, 
&&=~  R\,*\,L_*^{2n+1}\,*R\,|_{\bar{\theta}
\,\bar{\chi}\,\theta\,\chi\,}\, \\
\label{lrl2n1}
L\,*\,R\,*\,L_*^{2n+1}|_{\bar{\theta}
\,\bar{\chi}\,\theta\,\chi\,}\, 
&&=~L_*^{2n+2}\,*R\,|_{\bar{\theta}
\,\bar{\chi}\,\theta\,\chi\,}\, 
\eea
\item
Now, we proceed to establish few more identities, which 
will ultimately lead to the proof of the result given 
in eqn. (\ref{lrlrl}).
First, using the result for $\,L_*^{2m}\,$ given in 
eqn. (\ref{Lstara}) and also the result for 
$\,R\,*\,L_*^{2n}\,$ given
in eqn. (\ref{rl2n}), we can compute the following quantity: 
\bea  \label{Al2mrl2nI}
&&L_*^{2m}\,*\,(\,R \,*\,L_*^{2n}\,)|_{\bar{\theta}
\,\bar{\chi}\,\theta\,\chi\,}\, \xx
&&\!\!=~~
\Bigl(\,\frac{1}{4} \det~C\Bigr)^{m-1} \,F^{2m-2}
\left[ \,\,(-2m) \,
\Bigl(\,\frac{1}{4} \det~C\Bigr)^{n} \,F^{2n}
\,\bar{\psi}_L\,\bar{\psi}_R \,(-i\bar{F}) \right.\xx
&&\left.+\Bigl(\,\frac{1}{4} \det~C\Bigr)\,F^{2}(-i\bar{F})
\Bigl(\,\frac{1}{4} \det~C\Bigr)^{n-1} \,F^{2n-2}
\,(-2n\,\bar{\psi}_L\,\bar{\psi}_R \,) \right.\xx
&&\left.+\Bigl(\,\frac{1}{4} \det~C\Bigr)F^{2}
\Bigl(\,\frac{1}{4} \det~C\Bigr)^{n} \,F^{2n}
\partial_{\xi^-}\partial_{\zeta^-}\bar{A}
\,\right].
\eea
Independently, from the definition of $L_*^{2n}$ in
eqn. (\ref{Lstara}), one can also arrive at the following
relation:
\bea \label{Al2m2nr}
L_*^{2m+2n}\,*\,R |_{\bar{\theta}
\,\bar{\chi}\,\theta\,\chi\,}\,
&&=\left(\,\frac{1}{4} 
\det~C\right)^{m+n-1} \,F^{2(m+n)-2}
\,(-2m -2n) (-i\bar{F}\bar{\psi}_L\,\bar{\psi}_R) \xx
&&+ \left(\,\frac{1}{4} 
\det~C\right)^{m+n} \,F^{2(m+n)}
\partial_{\xi^-}\partial_{\zeta^-}\bar{A}.
\eea
Comparing eqns. (\ref{Al2mrl2nI}) and 
(\ref{Al2m2nr}), one can see that   
the following holds true:
\be \label{Al2mrl2n1}
L_*^{2m}\,*\,(\,R \,*\,L_*^{2n}\,) |_{\bar{\theta}
\,\bar{\chi}\,\theta\,\chi\,}\,
= L_*^{2m+2n} \,*\,R|_{\bar{\theta}
\,\bar{\chi}\,\theta\,\chi\,}\,.
\ee

\item
In a similar way, using the result derived
in eqn. (\ref{L1stara}) for $L_*^{2n+1}$,
one can explicitly deduce that:
\bea \label{Amn1r}
L_*^{2m+2n+1} \,*\,R |_{\bar{\theta}
\,\bar{\chi}\,\theta\,\chi\,}\, 
&&=~ 
~\left[\,\left(\frac{1}{4} \det~C\right)^{m+n}\, F^{2(m+n)-1}\,
\left(-2(m+n)\, i\,\bar{\psi}_L\,\bar{\psi}_R \,
\partial_{\xi^-}\partial_{\zeta^-}\bar{A}\right) \right.\xx
&&~~~~~~~~~~~~~~~ +
\left.F\left( i\,\bar{\psi}_L\,\partial_{\zeta^-}\psi_L
\,+\, i\,\bar{\psi}_R\,
\partial_{\xi^-}\psi_R \,-\, F\,{\bar F}\right)
\,\right].
\eea
On the other hand, using eqn. (\ref{Lstara}),
one can also show that:
\bea \label{Amnr1}
&&L_*^{2m}\,*\,(\,R \,*\,L_*^{2n+1}\,)|_{\bar{\theta}
\,\bar{\chi}\,\theta\,\chi\,}\, \xx
&&=~ 
~\left(\,\frac{1}{4} \det~C\right)^{n}\, F^{2n-1}\,
\left[-2n\, i\,\bar{\psi}_L\,\bar{\psi}_R \, 
L_*^{2m}\,*\,R \, 
+\,  F.L_*^{2m}\,*\,R \,*\,L\,\right]|_{\bar{\theta}
\,\bar{\chi}\,\theta\,\chi\,}\, \xx
&&=~\left(\,\frac{1}{4} \det~C\right)^{n}\, F^{2n-1}\,
\left[\,(-2\,i\,n\,)\,\bar{\psi}_L\,\bar{\psi}_R \,
\,\left(\frac{1}{4} \det~C\right)^{m}\, F^{2m}\,
\partial_{\xi^-}\partial_{\zeta^-}\bar{A} \right. \xx
&&\left.+\,F\left(\frac{1}{4} \det~C\right)^{m-1}\, F^{2m-1}\,
\Bigl\{ 
(-2\,m\,)\,\bar{\psi}_L\,\bar{\psi}_R \,i
\left(\frac{1}{4} \det~C\right) F 
\partial_{\xi^-}\partial_{\zeta^-}\bar{A} \right. \xx
&&\left.+ \left(\frac{1}{4} \det~C\right) F^2
\left( i\,\bar{\psi}_L\,\partial_{\zeta^-}\psi_L
\,+\, i\,\bar{\psi}_R\,
\partial_{\xi^-}\psi_R \,-\, F\,{\bar F}\right)
\Bigr\}
\,\right].
\eea
Thus, comparing 
eqns. (\ref{Amn1r})  and (\ref{Amnr1})we have:
\be
L_*^{2m}\,*\,(\,R \,*\,L_*^{2n+1}\,)|_{\bar{\theta}
\,\bar{\chi}\,\theta\,\chi\,}\, 
= L_*^{2m+2n+1} \,*\,R|_{\bar{\theta}
\,\bar{\chi}\,\theta\,\chi\,}\,.
\ee

\item
Now, using the definition of $L_*^{2n}$ given in 
eqn. (\ref{Lstara}), one can deduce that:
\bea \label{Al2mn1r}
&&L_*^{2(m+n+1)}\,*R|_{\bar{\theta}
\,\bar{\chi}\,\theta\,\chi\,}\, \xx
&&=~
\left(\frac{1}{4} \det~C\right)^{m+n}\, F^{2(m+n)}\,
\left[-2(m+n+1)\, i\,\bar{\psi}_L\,
\bar{\psi}_R \,(-i\bar{F}) \right. \xx
&& \left.
\,+ \left(\frac{1}{4} \det~C\right) F^2 \,
\partial_{\xi^-}\partial_{\zeta^-}\bar{A}
\right].
\eea
On the other hand, using the definition of 
$L_*^{2n+1}$ given in eqn. (\ref{L1stara}), 
one has:
\bea \label{Al2nr2m}
&&L_*^{2m+1}\,*R\,*L_*^{2n+1}\,|_{\bar{\theta}
\,\bar{\chi}\,\theta\,\chi\,}\, \xx
&&=~ \left(\frac{1}{4} \det~C\right)^{m+n}\, 
F^{2(m+n)-2}\,
\left[-2m\, i\,\bar{\psi}_L\,\bar{\psi}_R 
\,F\,R\,*\,L \right. \xx
&&~~~~\left.-2n\, i\,\bar{\psi}_L\,
\bar{\psi}_R \,F\,L\,*\,R
+ F^2 \,L\,*\,R*\,L
\right] |_{\bar{\theta}
\,\bar{\chi}\,\theta\,\chi\,}\,\xx
&&=~ \left(\frac{1}{4}\det~C\right)^{m+n}\, 
F^{2(m+n)-2}\,
\left[
-2m\, i\,\bar{\psi}_L\,\bar{\psi}_R
F (-F\bar{F}) -2n\, i\,\bar{\psi}_L\,\bar{\psi}_R
F (-F\bar{F}) \right. \xx
&&\left.+ F^2 
\Bigl\{ \left(\frac{1}{4} \det~C\right)
F^2 \partial_{\xi^-}\partial_{\zeta^-}\bar{A} 
+ 2i \bar{\psi}_L\,\bar{\psi}_R  \bar{F} 
\Bigr\}
\right].
\eea
Comparing eqns. (\ref{Al2mn1r}) and (\ref{Al2nr2m}), one
can see that:
\be
L_*^{2n+1}\,*R\,*L_*^{2n+1}\,|_{\bar{\theta}
\,\bar{\chi}\,\theta\,\chi\,}\,
= L_*^{2(m+n+1)}\,*R|_{\bar{\theta}
\,\bar{\chi}\,\theta\,\chi\,}\,.
\ee

\item
To show that:
\be \label{Aidentity1}
L_*^{2m}\,*R\,*L_*^{2n}\,*\,R\,*\,L_*^{2p}\,
|_{\bar{\theta}
\,\bar{\chi}\,\theta\,\chi\,}\,
~=~ L_*^{2(m+n+p)}\,*R_*^2|_{\bar{\theta}
\,\bar{\chi}\,\theta\,\chi\,}\,,
\ee
we directly calculate the following from 
the definitions of $\,L_*^{2p}\,$ and $R$ given
in eqns. (\ref{Lstara}) and (\ref{Ra}):
\bea \label{Alleven}
&&L_*^{2m}\,*R\,*L_*^{2n}\,*\,R\,*\,L_*^{2p}\,
|_{\bar{\theta}
\,\bar{\chi}\,\theta\,\chi\,}\,\xx
&&=~
\left(\,\frac{1}{4} 
\det~C\right)^{m+n+p-3} \,F^{2(m+n+p)-6}
\,\left[-2m \,(\theta \chi)\:
\bar{\psi}_L\,\bar{\psi}_R \,+\,
\left(\frac{1}{4}\det~C\right)\, 
F^2\, \right]\,*\,R\, \xx
&&\times \:
*\,\left[-2n \,(\theta \chi)\:
\bar{\psi}_L\,\bar{\psi}_R \, +\,  
\left(\frac{1}{4}\det~C\right)\, F^2\, \right]\,*\,R\,*
\,\left[-2p \,(\theta \chi)\:
\bar{\psi}_L\,\bar{\psi}_R \, +\,  
\left(\frac{1}{4}\det~C\right)\, F^2\, \right] \xx
&&=~
\left(\,\frac{1}{4} 
\det~C\right)^{m+n+p-3} \,F^{2(m+n+p)-6}
\,\left[-2(m+n+p)  \left(\,\frac{1}{4} 
\det~C\,F^2\right)^2 \right. \xx
&& \times \; \left.
\bar{\psi}_L\,\bar{\psi}_R \, 
(-2\psi_L\,\psi_R) \, 
+ 2\left(\,\frac{1}{4} 
\det~C\,F^2\right)^3 \,
\partial_{\xi^-}\partial_{\zeta^-}\bar{A}
\right].
\eea
On the other hand, one can also calculate the following
directly from eqn. (\ref{L1stara}):
\bea \label{Al2m2n2pr}
&&L_*^{2(m+n+p)}\,*\,R_*^2|_{\bar{\theta}
\,\bar{\chi}\,\theta\,\chi\,}\,   \xx
&&=~
\left(\,\frac{1}{4} 
\det~C\right)^{m+n+p-1} \,F^{2(m+n+p)-2}
\,\left[-2(m+n+p)\bar{\psi}_L\,\bar{\psi}_R \, 
(-2\psi_L\,\psi_R) \, \right. \xx
&&+\left. 2\left(\,\frac{1}{4} 
\det~C\,F^2\right) \,
\partial_{\xi^-}\partial_{\zeta^-}\bar{A}
\right].
\eea

\item
The following identity can be shown in an identical
fashion:
\be 
L_*^{2m+1}\,*R\,*L_*^{2n}\,*\,R\,*\,L_*^{2p}\,
|_{\bar{\theta}
\,\bar{\chi}\,\theta\,\chi\,}\,
~=~ L_*^{2(m+n+p)+1}\,*R_*^2|_{\bar{\theta}
\,\bar{\chi}\,\theta\,\chi\,}\,.
\ee

\item
Next, using the definitions 
of $\,L_*^{2p}\,$, $L_*^{2n+1}$ and $R$  
given in eqn. (\ref{L1stara}), 
(\ref{Lstara}) and (\ref{Ra})
one can directly calculate the following:
\bea \label{1even2oddI}
&&L_*^{2m+1}\,*R\,*L_*^{2n+1}\,*\,R\,*\,L_*^{2p}\,
|_{\bar{\theta}
\,\bar{\chi}\,\theta\,\chi\,}\,\xx
&&=~
\left(\,\frac{1}{4} 
\det~C\right)^{m} \,F^{2m-1}
\,\left[-2im \,
\bar{\psi}_L\,\bar{\psi}_R \,+
F.\,L \right]\,*\,R\,  \xx
&&~*\,\left(\,\frac{1}{4} 
\det~C\right)^{n} \,F^{2n-1}
\,\left[-2in \,
\bar{\psi}_L\,\bar{\psi}_R \,+
F.\,L \right]\,*\,R\, \xx
&&~~\times * \,\left(\,\frac{1}{4} 
\det~C\right)^{p-1} \,F^{2p-2}
\,\left[-2p \,(\theta\chi)
\bar{\psi}_L\,\bar{\psi}_R \,+
\left(\,\frac{1}{4} 
\det~C\right)\,F^2\, \right]\,  \xx
&&=~
\left(\,\frac{1}{4} 
\det~C\right)^{m+n+p-1} \,F^{2(m+n+p)-4}
\:\left[\:-2im \,
\bar{\psi}_L\,\bar{\psi}_R \, 
\left(\,\frac{1}{4} 
\det~C\right) F^3 \, R\,*\,L\,*\,R \,-\, \right. \xx
&&\left.\:-2in \,
\bar{\psi}_L\,\bar{\psi}_R \, 
\left(\,\frac{1}{4} 
\det~C\right) F^3 \, L\,*\,R\,*\,R \:
+\: F^2\,L\,*\, R\,*\,L\,*\,R \, (-2p(\theta\chi)) \right.\xx
&&~\left.+ \left(\,\frac{1}{4} \det~C\right) F^2
L\,*\, R\,*\,L\,*\,R \, 
\:\right]|_{\bar{\theta}
\,\bar{\chi}\,\theta\,\chi\,}\, \xx
&&=~
\left(\,\frac{1}{4} 
\det~C\right)^{m+n+p-1} \,F^{2(m+n+p)-4}
\:\left[\:-2im \,
\bar{\psi}_L\,\bar{\psi}_R \,
\left(\,\frac{1}{4} 
\det~C\right) F^3 \,(2i)\,\psi_L\,\psi_R \,F\right. \xx
&&\left.\:-2in \,
\bar{\psi}_L\,\bar{\psi}_R \, (2i) \,\psi_L\,\psi_R \,F
\left(\,\frac{1}{4} 
\det~C\right) F^3 \,
+ F^2\,(-2.\left(\,\frac{1}{4} 
\det~C\right) F^2\,\psi_L\,\psi_R) (-2p) \right. \xx
&&\left.+ \left(\,\frac{1}{4} \det~C\right)\, F^2\,
2\left\{\left(\,
\frac{1}{4} \det~C\right) F^2 
\partial_{\xi^-}\bar{A}
\partial_{\zeta^-}\bar{A} + 2
\bar{\psi}_L\,\bar{\psi}_R \,\psi_L\,\psi_R
\right\} \:\right].
\eea
In deriving the above identity, we have made use of the 
following results:
\bea
L\,*\,R\,*\,L\,*\,R \,|_{\bar{\theta}
\,\bar{\chi}\,\theta\,\chi\,}\,
&&=~~ R_*^2\,*\,L_*^2\,|_{\bar{\theta}
\,\bar{\chi}\,\theta\,\chi\,}\, \xx
&&=~~2\left\{\left(\,
\frac{1}{4} \det~C\right) F^2 
\partial_{\xi^-}\bar{A}
\partial_{\zeta^-}\bar{A} + 2
\bar{\psi}_L\,\bar{\psi}_R \,\psi_L\,\psi_R
\right\} ,\xx
L\,*\,R\,*\,L\,*\,R \,|_{\bar{\theta}
\,\bar{\chi}\,}\,
&&=~~-2.\left(\,\frac{1}{4} 
\det~C\right) F^2\,\psi_L\,\psi_R
\eea
Thus, one ends up with:
\bea \label{1even2oddII}
&&L_*^{2m+1}\,*R\,*L_*^{2n+1}\,*\,R\,*\,L_*^{2p}\,
|_{\bar{\theta}
\,\bar{\chi}\,\theta\,\chi\,}\,\xx
&&=~\left(\,\frac{1}{4} 
\det~C\right)^{m+n+p} \,F^{2(m+n+p)}
\left[4(m+n+p+1)\,
\bar{\psi}_L\,\bar{\psi}_R \,\psi_L\,\psi_R \right.\xx
&&\left.+ \,\left(\,\frac{1}{4} 
\det~C\right) \, F^2\,2\,
\partial_{\xi^-}\partial_{\zeta^-}\bar{A}
\right]|_{\bar{\theta}
\,\bar{\chi}\,\theta\,\chi\,} \xx
&&=~L_*^{2(m+n+p)+2}\,*R^2\,|_{\bar{\theta}
\,\bar{\chi}\,\theta\,\chi\,} \,.
\eea

\item
One can now go on and rigorously show that 
the following identity holds true as well
(when restricted to terms with coefficient
${\bar{\theta}\,\bar{\chi}\,(\theta\,\chi)\,}$):
\bea \label{identity2}
L_*^{2m+1}\,*R\,*L_*^{2n+1}\,*\,R\,*\,L_*^{2p+1}\,
|_{\bar{\theta}
\,\bar{\chi}\,\theta\,\chi\,}\,
&&\!\!=~~ L_*^{2(m+n+p)+3}\,*R_*^2|_{\bar{\theta}
\,\bar{\chi}\,\theta\,\chi\,}\,, 
\eea

In this manner, one finally arrives at the 
most general result that takes the form:
\be
L_*^{m}\,*R\,*L_*^{n}\,*\,R\,*\,L_*^{p}\,
|_{\bar{\theta}\,\bar{\chi}\,\theta\,\chi\,}\,
= L_*^{(m+n+p)}\,*R_*^2\,|_{\bar{\theta}
\,\bar{\chi}\,\theta\,\chi\,}\,,
\ee
where $m,n$ and $p$ are all integers.
\end{enumerate}


\end{document}